\documentclass[useAMS,usenatbib]{mnras}
\pdfoutput=1 
\usepackage{amsmath,amstext}
\usepackage[T1]{fontenc}
\usepackage{graphicx}
\usepackage[dvipsnames]{xcolor}
\usepackage[hang]{subfigure}
\usepackage{appendix}
\usepackage{array,multirow}
\usepackage{amssymb}
\usepackage{hyperref}

\usepackage{booktabs}
\usepackage{multirow}

\usepackage[normalem]{ulem}

\newcommand{\txn}[1]{\textnormal{#1}}

\newcommand{\beagle}{\textsc{beagle}}
\newcommand{\cloudy}{\textsc{cloudy}}
\newcommand{\mappings}{\textsc{mappings}}

\newcommand{\multinest}{\textsc{MultiNest}}
\newcommand{\pyspeclines}{\textsc{PySpecLines}}
\newcommand{\pyspeckit}{\textsc{PySpecKit}}
\newcommand{\HIIChimistry}{\textsc{Hii-Chi-mistry}}
\newcommand{\NebularBayes}{\textsc{NebularBayes}}

\newcommand{\M}{\hbox{$\txn{M}$}}
\newcommand{\Mstar}{\hbox{$M_{\star}$}}
\newcommand{\MstarInLog}{\hbox{$\txn{M}$}}

\newcommand{\Msun}{\hbox{$\M_{\sun}$}}

\newcommand{\Mbh}{\hbox{$M_{\textsc{bh}}$}}

\newcommand{\Mtot}{\hbox{$\M_\txn{tot}$}}
\newcommand{\MtotInLog}{\hbox{$\txn{M}_\txn{tot}$}}
\newcommand{\yr}{\hbox{$\txn{yr}$}}

\renewcommand{\t}{\hbox{$t$}}

\newcommand{\sfr}{\hbox{${\Psi}$}}
\newcommand{\sfrInLog}{\hbox{${\psi}$}}

\newcommand{\logOH}{\hbox{$12 + \log (\txn{O}/\txn{H})$}}
\newcommand{\Z}{\hbox{$\txn{Z}$}}
\newcommand{\Zism}{\hbox{$\Z_\textsc{ism}$}}
\newcommand{\Zhii}{\hbox{$\Z_\txn{gas}^\textsc{hii}$}}

\newcommand{\Zsun}{\hbox{$\Z_{\odot}$}}

\newcommand{\Epeak}{\hbox{$\txn{E}_\txn{peak}$}}

\newcommand{\HII}{\mbox{H\,{\sc ii}}}
\newcommand{\HI}{\mbox{H\,{\sc i}}}
\newcommand{\nH}{\hbox{$n_\textsc{h}$}}

\newcommand{\logUs}{\hbox{$\log U_\textsc{s}$}}

\newcommand{\xid}{\hbox{$\xi_\txn{d}$}}
\newcommand{\COsol}{\hbox{(C/O)$_\odot$}}

\newcommand{\tauV}{\hbox{$\hat{\tau}_\textsc{v}$}}
\newcommand{\mud}{\hbox{$\mu$}}

\newcommand{\Lacc}{\hbox{$L_\txn{acc}$}}
\newcommand{\PLalpha}{\hbox{$\alpha_\textsc{pl}$}}

\newcommand{\logUsAGN}{\hbox{$\log U_\textsc{s}^\textsc{nlr}$}}
\newcommand{\logUinAGN}{\hbox{$\log U_\txn{in}^\textsc{nlr}$}}

\newcommand{\UsAGN}{\hbox{$U_\txn{s}^\textsc{nlr}$}}
\newcommand{\xidAGN}{\hbox{$\xi_\txn{d}^\textsc{nlr}$}}
\newcommand{\ZAGN}{\hbox{$\Z_\txn{gas}^\textsc{nlr}$}}
\newcommand{\nHAGN}{\hbox{$n_\textsc{h}^\textsc{nlr}$}}

\newcommand{\oii}{\relax \ifmmode {\mbox O\,{\scshape ii}}\else O\,{\scshape ii}\fi}
\newcommand{\oiii}{\relax \ifmmode {\mbox O\,{\scshape iii}}\else O\,{\scshape iii}\fi}
\newcommand{\OIIIauroral}{\hbox{[O{\sc iii}]$\lambda4363$}}
\newcommand{\Halpha}{\hbox{H$\alpha$}}
\newcommand{\Hbeta}{\hbox{H$\beta$}}
\newcommand{\OI}{\hbox{[O\,{\sc i}]$\lambda6300$}}
\newcommand{\NII}{\hbox{[N\,{\sc ii}]$\lambda6584$}}
\newcommand{\SII}{\hbox{[S\,{\sc ii}]$\lambda6717,\lambda6731$}}
\newcommand{\OII}{\hbox{[O\,{\sc ii}]$\lambda3726,\lambda3729$}}
\newcommand{\OIII}{\hbox{[O\,{\sc iii}]$\lambda5007$}}
\newcommand{\OIIIall}{\hbox{[O\,{\sc iii}]$\lambda4959,\lambda5007$}}
\newcommand{\HeIIUV}{\hbox{[He\,{\sc ii}]$\lambda1640$}}
\newcommand{\HeII}{\hbox{[He\,{\sc ii}]$\lambda4686$}}

\newcommand{\NV}{\hbox{[N\,{\sc v}]$\lambda1240$}}
\newcommand{\HeI}{\hbox{He\,{\sc i}$\lambda5876$}}
\newcommand{\NeIII}{\hbox{[N\,{\sc iii}]$\lambda3869$}}

\newcommand{\Lsun}{\hbox{\,${\rm L}_\odot$}}
\newcommand{\ergscm}{\hbox{$\textrm{erg}\,\textrm{s}^{-1}\,\textrm{cm}^{-2}$}}

\newcolumntype{L}[1]{>{\raggedright\let\newline\\\arraybackslash\hspace{0pt}}m{#1}}
\newcolumntype{C}[1]{>{\centering\let\newline\\\arraybackslash\hspace{0pt}}m{#1}}
\newcolumntype{R}[1]{>{\raggedleft\let\newline\\\arraybackslash\hspace{0pt}}m{#1}}

\title[\textsc{beagle-agn}]{\beagle-\textsc{agn} I: Simultaneous constraints on the properties of gas in star-forming and AGN narrow-line regions in galaxies}
\author[Vidal-Garc\'ia et al.]{A. Vidal-Garc\'ia$^{1,2}$\thanks{Email: a.vidal@oan.es},
A. Plat$^{3}$\thanks{Email: plat@iap.fr}, E. Curtis-Lake$^{4}$\thanks{Email: e.curtis-lake@herts.ac.uk}, A. Feltre$^{5}$, M. Hirschmann$^{6,7}$, J. Chevallard$^{8}$ 
\newauthor  and
S. Charlot$^{9}$\\
$^{1}$ Observatorio Astronómico Nacional, C/ Alfonso XII 3, 28014 Madrid, Spain\\
$^{2}$ École Normale Supérieure, CNRS, UMR 8023, Université PSL, Sorbonne Université, Université de Paris, F-75005 Paris, France\\
$^{3}$ Steward Observatory, 933 N. Cherry Avenue, University of Arizona, Tucson, AZ 85721, USA \\
$^{4}$ Centre for Astrophysics Research, Department of Physics, Astronomy and Mathematics, University of Hertfordshire, Hatfield AL10 \\9AB, UK \\
${5}$ INAF  -  Osservatorio  di  Astrofisica  e  Scienza  dello  Spazio  di  Bologna,  Via  P.  Gobetti  93/3,  40129  Bologna,  Italy\\
$^{6}$ Institute of Physics, GalSpec Laboratory, Ecole Polytechnique Federale de Lausanne, Observatoire de Sauverny, Chemin Pegasi 51, \\ 1290 Versoix, Switzerland \\
$^{7}$ INAF, Osservatorio Astronomico di Trieste, Via G. B. Tiepolo 11, 34131 Trieste, Italy \\
$^{8}$ Sub-department of Astrophysics, Department of Physics, University of Oxford, Denys Wilkinson Building, Keble Road,
Oxford OX1 \\3RH, UK \\
$^{9}$ Sorbonne Universit\'{e}, CNRS, UMR7095, Institut d'Astrophysique de Paris, F-75014, Paris, France}

\begin{document}

\date{}

\pagerange{\pageref{firstpage}--\pageref{lastpage}} \pubyear{2018}

\maketitle

\label{firstpage}
\def\Msun{\hbox{$\rm\thinspace M_{\odot}$}}

\begin{abstract}
We present the addition of nebular emission from the narrow-line regions (NLR) surrounding active galactic nuclei (AGN) to \beagle\ (BayEsian Analysis of GaLaxy sEds).
Using a set of idealised spectra, we fit to a set of observables (emission-line ratios and fluxes) and test the retrieval of different physical parameters.  We find that fitting to standard diagnostic-line ratios from \cite{BPT} plus \OII/\OIII, \Hbeta/\Halpha, \OI/\OII\ and \Halpha\ flux, degeneracies remain between dust-to-metal mass ratio (\xidAGN) and ionisation parameter (\UsAGN) in the NLR gas, and between slope of the ionizing radiation (\PLalpha, characterising the emission from the accretion disc around the central black hole) and total accretion-disc luminosity (\Lacc).  Since these degeneracies bias the retrieval of other parameters even at maximal signal-to-noise ratio (S/N), without additional observables, we suggest fixing \PLalpha\ and dust-to-metal mass ratios in both NLR and \HII\ regions.  We explore the S/N in \Hbeta\ required for un-biased estimates of physical parameters, finding that S/N$(\Hbeta)\sim10$ is sufficient to identify a NLR contribution, but that higher S/N is required for un-biased parameter retrieval ($\sim20$ for NLR-dominated systems, $\sim30$ for objects with approximately-equal \Hbeta\ contributions from NLR and \HII\ regions). We also compare the predictions of our models for different line ratios to previously-published models and data. By adding \HeII-line measurements to a set of published line fluxes for a sample of 463 AGN NLR, we show that our models with $-4<\logUsAGN<-1.5$ can account for the full range of observed AGN properties in the local Universe.  

\end{abstract}


\newpage

\section{Introduction}


In searching for obscured (termed type-2) Active Galactic Nuclei (AGNs), astronomers  often exploit `BPT' diagrams, which are named after the founding paper by \cite{BPT}.  The scheme proposed in \cite{BPT}, which employs ratios between strong emission line fluxes, was later updated by \cite{Veilleux1987}, who settled on three diagnostic diagrams: \OIII/\Hbeta\ versus \SII/\Halpha, \OIII/\Hbeta\ versus \NII/\Halpha\ and \OIII/\Hbeta\ versus \OI/\Halpha.  These diagrams provide clear separation of the different ionizing sources powering the line emission.
However, at low metallicities, the regions within the diagnostic diagrams occupied by star-formation and AGN-driven line emission increasingly overlap, muddying the classification \citep{Groves2006, Feltre2016, Hirschmann2019}. Therefore, at low masses, or higher redshifts, the standard BPT diagrams will be less helpful to distinguish AGN from star-forming activity.
It is therefore important to develop new tools to probe the impact of AGNs on galaxy evolution with the new era of large spectroscopic surveys at high redshifts (e.g. with the \textit{James Webb Space Telescope}).

Beyond simple classification of galaxy-wide star-formation and AGN-driven line emission, there has been significant progress in deriving physical properties of the star-forming regions themselves  \citep[e.g.][and references therein]{Kewley2019, MaiolinoReview}. However, progress on the side of AGN gas properties is more limited.   The first emission-line calibrations to derive oxygen abundances of type-2 AGNs were supplied by \cite{Storchi-Bergmann1998}.  The subsequent efforts have proceeded along three avenues: 
further research of observables to derive physical properties (e.g. from rest-frame UV lines, \citealt{Dors2014}; or the rest-frame optical, \citealt{Castro2017}); comparison of emission-line ratios with photoionization models, often searching for a minimum $\chi^2$ solution \citep[e.g. for rest-frame UV lines,][]{Nagao2006,Matsuoka2009,Matsuoka2018}; and full Bayesian parameter estimation \citep{Perez-Montero2019,Thomas2018,Mignoli2019}. 

\cite{Dors2020} used a large sample of type-2 AGNs selected from the Sloan Digital Sky Survey \citep[SDSS,][]{York2000} to provide the first comparison of methods for deriving oxygen abundances of type-2 AGNs.  
These methods include the calibrations of \cite{Storchi-Bergmann1998}, \cite{Castro2017} and the Bayesian-style method employed by the \HIIChimistry\ code \citep{Perez-Montero2019}.  They find fairly poor agreement between the different methods as well as no significant trend between a galaxy's stellar mass and the metallicity of the narrow-line region (NLR) surrounding the AGN.

\cite{Dors2020} also test a `direct' method, that estimates the electron temperature ($T_\txn{e}$) within the high-ionization zone using the $R_{\mathrm O3}=[\mathrm O\textsc{iii}]\lambda4959\AA,\lambda5007\AA/\lambda4363\AA$ ratio.  They find that their implemented `direct' method significantly under-estimates the derived metallicities compared to the other methods.  In \cite{Dors2015}, they inferred that the `direct' metallicities were significantly lower than extrapolated radial metallicity gradients derived from gas-phase abundances.  \cite{Dors2020b} subsequently identify the cause of the discrepancy between various calibrations and the `direct' method. They update the `direct' method and demonstrate that the previous discrepancies with other calibrations are much reduced. Further to this, \cite{Dors2021} provide the first strong-line calibration against `direct' metallicity estimates for AGNs.

Despite large discrepancies in different oxygen abundance estimates, there have been efforts to characterise the type-2 AGN population by studying line emission.  The first studies investigating the statistics of gas metallicity in the narrow-line region of AGNs mostly indicated a lack of evolution with redshift \citep{Matsuoka2009, Dors2014}, although they did reveal a luminosity-metallicity relation, where less luminous AGNs (characterised by the \HeIIUV\ luminosity in both these studies) display lower metallicities than their bright counterparts. In a later study, \cite{Mignoli2019} investigated the properties of the NLR gas of a sample of Type-2 AGNs selected in a homogeneous way
and found significant evolution with redshift.  This study employed a wider selection of lines, as well as updated photoionization models,
which better reproduce the \NV\ line without resorting to the very high metallicities evoked in previous studies of quasars \citep[e.g.,][]{Hamann1993,Dietrich2003, Nagao2006b} and narrow-line Seyfert-I galaxies \citep{Shemmer2002}. Recently, \cite{Nascimento2022} used strong-line calibrations to study the NLR in type-2 AGNs within the MaNGA survey \citep[Mapping Nearby Galaxies at Apache Point Observatory;][]{Bundy2015}, finding that the central NLR region typically has lower metallicity than the surrounding HII regions. They posit that this may be due to accretion of metal-poor gas at the centre of these galaxies, which is feeding the central black hole.

One limitation of all the methods mentioned thus far is the lack of any accounting for \HII\ region contribution to the line emission.  \cite{Thomas2018} explicitly investigate the degree of \HII\ and NLR mixing in the line emission in a sample of SDSS galaxies using their Bayesian code, \NebularBayes.  They demonstrate that even in regions of the BPT diagram where AGNs are determined to be cleanly selected, the Balmer lines can have a significant contribution from \HII\ regions ionized by young stars \citep[see also ][]{Agostino2021}.  Further to this, \cite{Thomas2019} measure the mass-metallicity relation of type-2 AGNs finding a moderate increase in oxygen abundance with increasing stellar mass. 

With this paper, we present the incorporation of the \cite{Feltre2016} NLR models into \beagle\ (BayEsian Analysis of GaLaxy sEds), a tool to model and interpret galaxy spectral energy distributions \citep{Chevallard2016} (Section~\ref{section:modelling}).  This addition allows the mixing of line emission from young stellar birth clouds (Section~\ref{section:stars_hii}) with that from the NLR of type-2 AGNs (Section~\ref{section:agn_nlr}).  \beagle\ also self-consistently includes stellar emission and attenuation by interstellar dust (Section~\ref{section:dust}), which are not explicitly modelled in \NebularBayes.

In Section~\ref{section:retrieval}, we take a pedagogical approach to defining what parameters of our model can be constrained by fitting a given set of observables   (Section~\ref{section:fitting}) in idealised spectra  (Section~\ref{section:grid_description}).  Using such spectra, we quantify the S/N required to constrain \HII-region and NLR-gas parameters for different NLR contributions to the total \Hbeta\ flux of a galaxy  (Sections~\ref{section:highqual} and~\ref{section:influenceSN}).  With this work, we focus on rest-frame optical observables, though \beagle\ can also be used to study emission lines from the rest-frame UV. Finally, we compare the results of our model with those obtained using previously published methods in Section~\ref{section:discussion} and expand on the comparison work of \cite{Dors2020} by explaining how different methods will derive different oxygen abundances (Section~\ref{section:approachesNLR}). In particular, in Section~\ref{section:generalcomparison} we compare several emission-line ratios and several free parameters in the models and in a set of type-2 AGN observations. In order to explain different trends for the ionization parameter,  in Section~\ref{section:HeII} we present new measurements of the \HeII\ fluxes in a few hundred type-2 AGNs from DR7 SDSS and we compare the measured data to the fluxes predicted by the F16 models.  We also show the predictions of the fluxes for sulfur and nitrogen lines in Section~\ref{section:nitrogen}, and end this discussion
(Section~\ref{section:comparingDors}) by comparing the \cite{Dors2021} empirical $T_\txn{e}$-based \logOH\ calibration to the different model grids. The paper ends in Section~\ref{section:summary} with a summary of the main findings of this work. This is the first in a series of three papers, to be followed by a paper on fitting of a sample of type-2 AGNs with \beagle, and a study of the extent to which line emission from shocks and post-AGB stars may affect our inferences.

\section{Modelling the emission from stars, \HII\ regions and AGN NLRs in \beagle}\label{section:modelling}

\beagle\ \citep[hereafter CC16]{Chevallard2016}  is a Bayesian SED-fitting code, which allows efficient exploration of a wide grid of physical parameters affecting the light emitted by a galaxy. 
The code employs \multinest\ \citep[a Bayesian analysis tool based on the Nested Sampling algorithm of \citealt{Skilling2006}; see][]{Feroz2009} to sample from the posterior probability distributions of physical parameters.  We refer the reader to CC16 for more details.

A main feature of \beagle\ is the incorporation of physically consistent stellar continuum and nebular emission models, which trace the production of starlight and its transmission through the interstellar medium self-consistently.  This paper is concerned with the extension of \beagle\ to allow for the interpretation of mixed emission-line signatures of AGNs and stars.  In Section~\ref{section:stars_hii} below, we briefly review the modelling of the emission from stars and \HII\ regions in \beagle, while in Section~\ref{section:agn_nlr}, we describe our incorporation of the emission from AGN narrow-line regions.

\subsection{Emission from stars and \HII\ regions}
\label{section:stars_hii}

\beagle\ employs the latest version of the \citet{Bruzual2003} stellar population synthesis models \citep[as described in][]{Vidal-Garcia2016} to compute the emission from stars.

The line and continuum emission of \HII\ regions ionized by stars younger than 10\,Myr  is computed following the prescription of \citet[][see also \citealt{Charlot2001}]{Gutkin2016} for a grid of gas parameters: the metallicity of the \HII-region gas, \Zhii\ (noted \Zism\ in \citealt{Gutkin2016}); the ionization parameter, \logUs\footnote{Note that \logUs{} is defined as the ionization parameter at the Str\"omgrem radius, which differs from the volume-averaged ionization parameter, $<U>$ according to $<U> = 9/4\,$U$_\textsc{s}$ }; 
and the dust-to-metal mass ratio within the \HII\ regions, \xid.  Nitrogen abundances are scaled to oxygen abundances according to the formula:
\begin{equation}
    {\rm N/H}\simeq 0.41 \, {\rm O/H}\left[10^{-1.6} + 10^{(2.33 + \log \rm O/H)}\right]
    \label{eq:Nitrogen}\,,
\end{equation}
which well matches the relation between N/O and \logOH\ in observed \HII\ regions \citep[see equation 11 and figure 1 of][]{Gutkin2016}.


For simplicity, we fix here the carbon-to-oxygen ratio of the \HII\ regions to solar, $\COsol=0.44$, and the hydrogen density to $\nH=100\,$cm$^{-3}$. In the stellar population models, we fix the upper limit of the stellar initial mass function (IMF) to 100 \Msun.  An important feature of this grid is the coupling of stellar light to its transmission through the interstellar medium (ISM), via not only the nebular emission, but also the attenuation by dust in the \HII\ regions themselves (see Section~\ref{section:dust}).

\subsection{Emission from AGN narrow-line regions}
\label{section:agn_nlr}

We appeal to the \citet[hereafter F16]{Feltre2016} models of AGN-NLR emission to introduce simultaneous fitting of the physical properties of NLR and \HII\ regions with \beagle. The F16 NLR models were produced with 
\cloudy\ c13.03 
\citep[][the same version as used by \citealt{Gutkin2016} to produce the \HII-region models]{Ferland2013}.  Three important aspects must be considered to incorporate the treatment of NLR emission into \beagle: the shape of the AGN ionizing continuum; the properties of the NLR gas itself; and the integration of the F16 model grid with the \HII-region model grid within \beagle.  In the next paragraphs, we summarise the salient features of our approach to achieve these goals.

\begin{figure}
  \centering
  \includegraphics[width=3.5in]{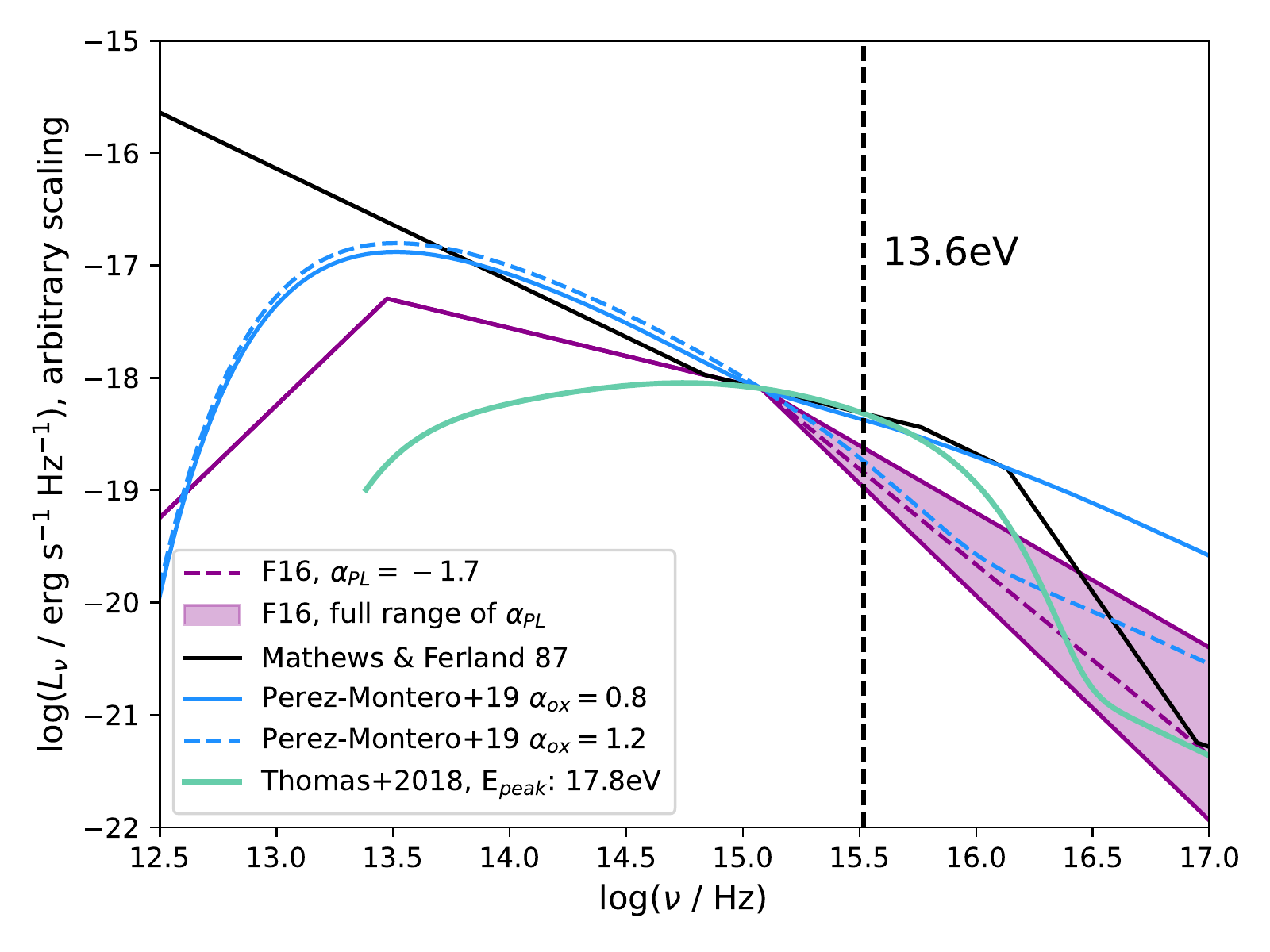}
  \caption{The incident radiation used to compute different NLR emission models.  The models of F16 are used in this work, and allow a range of slope (\PLalpha) of the ionizing portion of the spectrum. The F16 models use a simplified parametric form which describes the thermal emission from the accretion disc surrounding the central black hole.  We use a model grid that includes a range of spectral indices, \PLalpha, in the rest-frame UV to soft X-ray [$\log(\nu/\txn{Hz}) \gtrsim 15$], as indicated.  The shape is described in \protect\cite{Schartmann2005} and \protect\cite{Feltre2012}.  We also plot the incident radiation used in a number of different works for comparison, as specified in the legend.  We compare the emission line ratios of these different models to observations in Section \ref{section:discussion}.}
  \label{fig:accretion disc}
\end{figure}

In the F16 model, the radiation illuminating the NLR gas is the thermal emission from the accretion disc surrounding the central black hole.  For simplicity, this emission is described by a broken power law, as shown in Fig.~\ref{fig:accretion disc} (see also equation 5 of F16).  The strengths of emission lines emerging from the NLR, as well as the ratios between them, are primarily sensitive to the slope, \PLalpha, of the power law at high frequencies (short wavelengths).  F16 allow this parameter to vary in the range $-1.2<\PLalpha<-2.0$, which are generally used as ``standard'' values for the power-law slope \citep[e.g.][]{Groves2004}. 

This simplified description of the purely thermal emission from the accretion disc does not attempt to model the hard X-ray component thought to arise from the Comptonization of photons scattered in the hot corona surrounding the disc \citep{Haardt1991}.  A soft X-ray component is also often observed in AGNs, the exact nature of which is still debated (e.g., disc reflection, \citealt{Crummy2006}; or optically-thick Comptonized disc emission, which can explain the soft X-ray excess in narrow-line Seyfert-Is, \citealt{Done2012}).  In practice, the flexibility in \PLalpha\ allows the exploration of different contributions of soft X-rays to the incident ionizing radiation. 

Once the spectrum of the accretion-disc radiation reaching the NLR is defined, \cloudy\ can be used to model the transfer of this radiation through the NLR gas.  The F16 NLR grid employs the option of `open geometry', suitable for gas with low covering fraction.  The models are computed for a typical accretion luminosity (i.e., the integral of the broken power law in Fig.~\ref{fig:accretion disc}) of $\Lacc=10^{45}\,\ergscm$ (denoted $L_\txn{AGN}$ in F16). The version of the F16 model incorporated into \beagle\ includes further updates by \cite{Mignoli2019} which better account for the observed \NV/\HeIIUV\ emission-line ratios in a sample of high-redshift type-2 AGNs. Specifically, the inner radius of the gas is set to $r_\txn{in}\approx90\,\textrm{pc}$, and the internal micro-turbulence velocity to $v_\txn{micr}=100\,\textrm{km}\,\textrm{s}^{-1}$.  

To incorporate nebular emission from AGN NLR into \beagle, we use a grid of F16 models covering full ranges in \PLalpha\ and other NLR-gas parameters: the metallicity, noted \ZAGN; the ionization parameter, \logUsAGN; and the dust-to-metal mass ratio, \xidAGN. In this work, we fix the hydrogen density in the NLR to $\nHAGN= 1000$\,cm$^{-3}$, as typically measured from optical line doublets \citep{Osterbrock2006}. Furthermore, as for the \HII\ regions (Section~\ref{section:stars_hii}), we fix the C/O abundance ratio in the NLR gas to solar and use equation~(\ref{eq:Nitrogen}) to describe the dependence of N/H on O/H. Table~\ref{tab:physical_parameter_list} lists the full set of physical parameters pertaining to the \HII\ and NLR regions in our models.

\begin{table*}
\centering
  \caption{List of physical parameters used in this work to simulate and fit to (active and inactive) galaxy spectra. }
  \begin{tabular}{l|l}
  \hline
  Parameter & Description\\
  \hline
   $\MtotInLog/\Msun$ & Integrated SFH. \\
   \MstarInLog/\Msun & Stellar mass, including stellar remnants.\\
  $\sfrInLog/\Msun\mathrm{yr}^{-1}$ & Current star formation rate.\\
  $\Zhii/\Zsun$, $\ZAGN/\Zsun$  & Metallicity of gas in \HII\ regions and the NLR, respectively.\\   
  \tauV & Total $V$-band attenuation optical depth in the ISM.\\
  $\mud$ & Fraction of \tauV\ arising from dust in the diffuse ISM, fixed to 0.4.\\
  \logUs, \logUsAGN & Effective gas ionization parameter in \HII\ regions and the NLR, respectively.\\
  \xid, \xidAGN & Dust-to-metal mass ratio in \HII\ regions and the NLR, respectively.\\
  $\nH/\mathrm{cm}^{-3}$, $\nHAGN/\mathrm{cm}^{-3}$ & Hydrogen gas density in \HII\ regions (fixed to 100) and the NLR (fixed to 1000), respectively.\\
  (C/O)/(C/O)$_{\odot}$ & Carbon-to-oxygen abundance ratio in units of  (C/O)$_{\odot}=0.44$, fixed to unity.\\
  $m_{\mathrm{up}}/\Msun$ & Upper mass cutoff of the IMF, fixed to 100.\\
  \t/Gyr & Age of the oldest stars.\\
  \PLalpha & Slope of the ionizing radiation from thermal emission of the accretion disc around the central black hole.\\
  \Lacc & Integrated thermal emission from the accretion disc.\\
  \hline
\end{tabular}
\label{tab:physical_parameter_list}
\end{table*}

In combining the \HII-region and NLR model grids in \beagle, we allow for the normalisation of the NLR models to change. In practice, we achieve this by changing the parameter \Lacc, which controls the absolute luminosities of NLR emission lines (Table~\ref{tab:physical_parameter_list}). We note that this is compatible with the adoption of a fixed $\Lacc=10^{45}\,\ergscm$ in the photoionization calculations of F16.  Indeed, we find that, at fixed other parameters, the emission-line luminosities computed with \cloudy\ for values of \Lacc\ in the range from $10^{42}$ to $10^{48}\,\ergscm$ differ by only 0.1 to 6 per cent (depending on the considered line) from those obtained by simply scaling the luminosities of a model with $\Lacc=10^{45}\,\ergscm$. This is because we assume a fixed scaling relation between the inner radius of the ionized gas ($r_{\rm in}$) and  \Lacc, given by \Lacc$/(4 \pi r_{\rm in}^{2})\simeq10^2\,\ergscm$ \citep{Feltre2016}. \footnote{This negligible influence of \Lacc\ on model predictions at fixed other parameters (including \logUsAGN) arises from a degeneracy between \Lacc\ and the volume-filling factor of the gas entering the definition of the ionization parameter (see equation~4 of F16).}
We must also account for the inhomogeneous distribution of the narrow-line emitting gas around the central AGN. This can be achieved by multiplying \Lacc\ by a gas covering fraction, whose recovery from observables will however be degenerate with \Lacc. To avoid introducing this extra degeneracy when fitting to data with \beagle, we fix the covering fraction of NLR gas to 10 per cent, which is within the range of values (2-20$\%$) obtained for the covering fraction of the NLR gas of the Palomar-Green quasar sample \citep{Baskin2005}.


Since we are modelling type-2 AGNs, we are focused on modelling the narrow line-emitting region.  Continuum emission from the accretion disc itself is assumed to be completely obscured by the dust surrounding the AGN centre, often referred to as the AGN torus.  Any non-ionizing continuum light (of the form in Fig.~\ref{fig:accretion disc}) that is reflected from the NLR is included in the models, 
however, its contribution to most SEDs is negligible relative to the stellar continuum in practice. 
For simplicity, we assume that the radiation by the accretion disc and surrounding NLR gas are spherically symmetric.  This is a reasonable assumption when teamed with the covering fraction, but does not take account of scattering or absorption of the accretion disc radiation by the dust in the torus, nor of the anisotropic nature of the emission from the accretion disc itself.



\subsection{Attenuation by dust}
\label{section:dust}

In this work, we adopt the \citet[hereafter CF00]{Charlot2000} two-component dust model.  This model accounts for the enhanced dust content in stellar birth clouds compared to the diffuse ISM.  The dust within birth clouds is split between the inner \HII\ and outer \HI\ regions, with the dust in \HII\ regions being self-consistently accounted for in the nebular emission grid of \cite{Gutkin2016}.  As of \beagle\ v0.27.1, this dust component is accounted for self-consistently within the implementation of the CF00 model, as described in \citet{Curtis-Lake2021}.

When combining NLR emission with the emission from stars we account for the dust both within the NLR and the diffuse ISM. As for the \HII\ region models, dust is included in the F16 \cloudy\ models of the NLR. Hence, within \beagle, in the framework of the CF00 dust model, light emitted from the NLR is subject only to further attenuation by dust in the diffuse ISM.

\section{Retrieval of parameters from idealised models}
\label{section:retrieval}

\subsection{Idealised models of active galaxies}
\label{section:grid_description}


\begin{table}
\centering
  \caption{Properties of the idealised models of active galaxies at $z=0$ and $z=2$ used to test parameter retrieval with \beagle.}
  \begin{tabular}{l|r|r}
  \hline
  Parameter & $z = 0$ & $z = 2$  \\
  \hline
  $\log(\MstarInLog/\Msun)$ & 10.5 & 10.5 \\
  $\sfrInLog/\Msun\mathrm{yr}^{-1}$ & 0.4 & 1.3 \\
  \Zhii & 0.014 & 0.007 \\
  \logUs & $-3.6$ & $-3.0$ \\
  $\nH/\txn{cm}^{-3}$ & 100 & 100 \\
  \xid & 0.3 & 0.3 \\
  \t/Gyr & 10 & 1 \\
  \ZAGN & 0.030 & 0.012 \\
  \logUsAGN & $-3.5$ & $-2.75$ \\
   $\nHAGN/\txn{cm}^{-3}$ & 1000 & 1000 \\
  \xidAGN & 0.3 & 0.3 \\
  \tauV & 0.3 & 0.3 \\
  $\mu$ & 0.4 & 0.4 \\
  $\PLalpha$ & $-1.7$ & $-1.7$ \\
 \hline
\end{tabular}
\label{tab:grid_table}
\end{table}

\begin{figure*}
  \centering
  \includegraphics[width=7.0in]{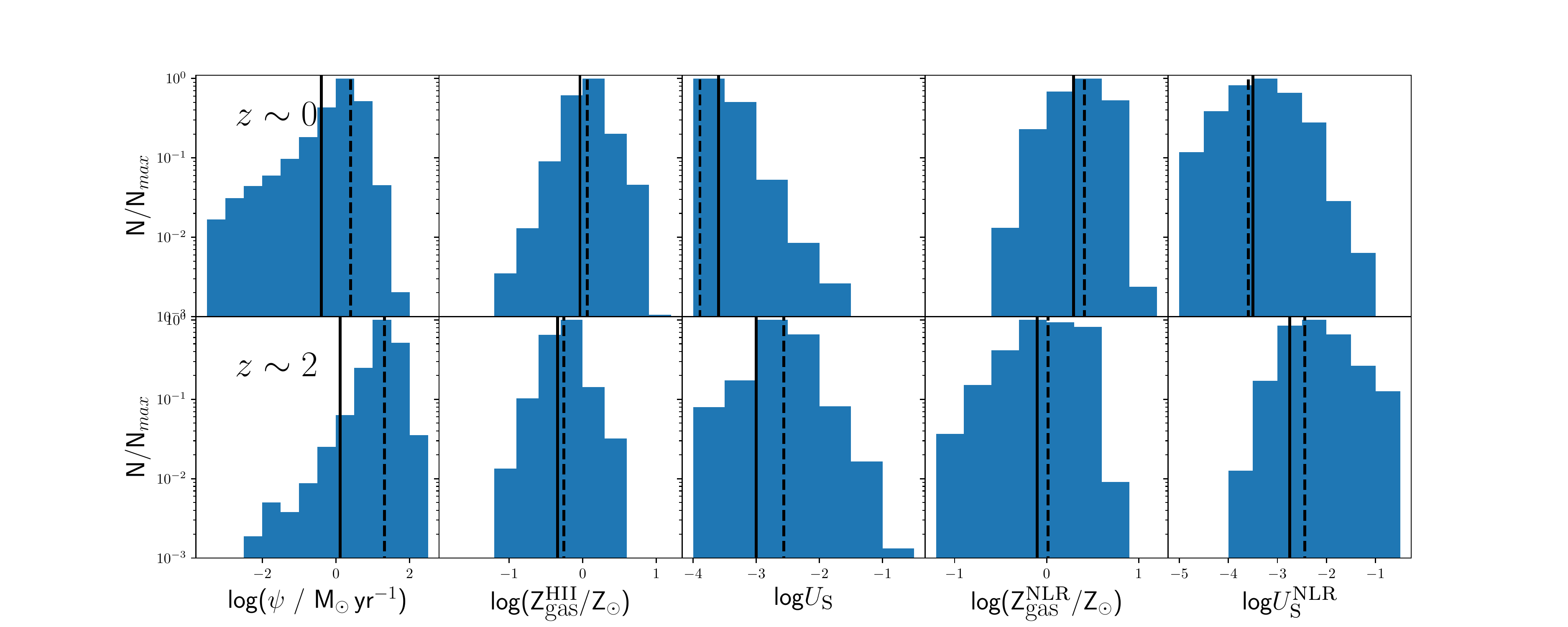}
  \caption{Histograms of the physical parameter space covered by galaxies in IllustrisTNG-100 at z=0 (top row) and z=2 (bottom row). Dashed vertical lines indicate the mean value of each quantity, while the solid black lines indicate the values used.}
  \label{fig:histo_param_input}
\end{figure*}

\begin{figure*}
  \centering
  \includegraphics[width=7.0in]{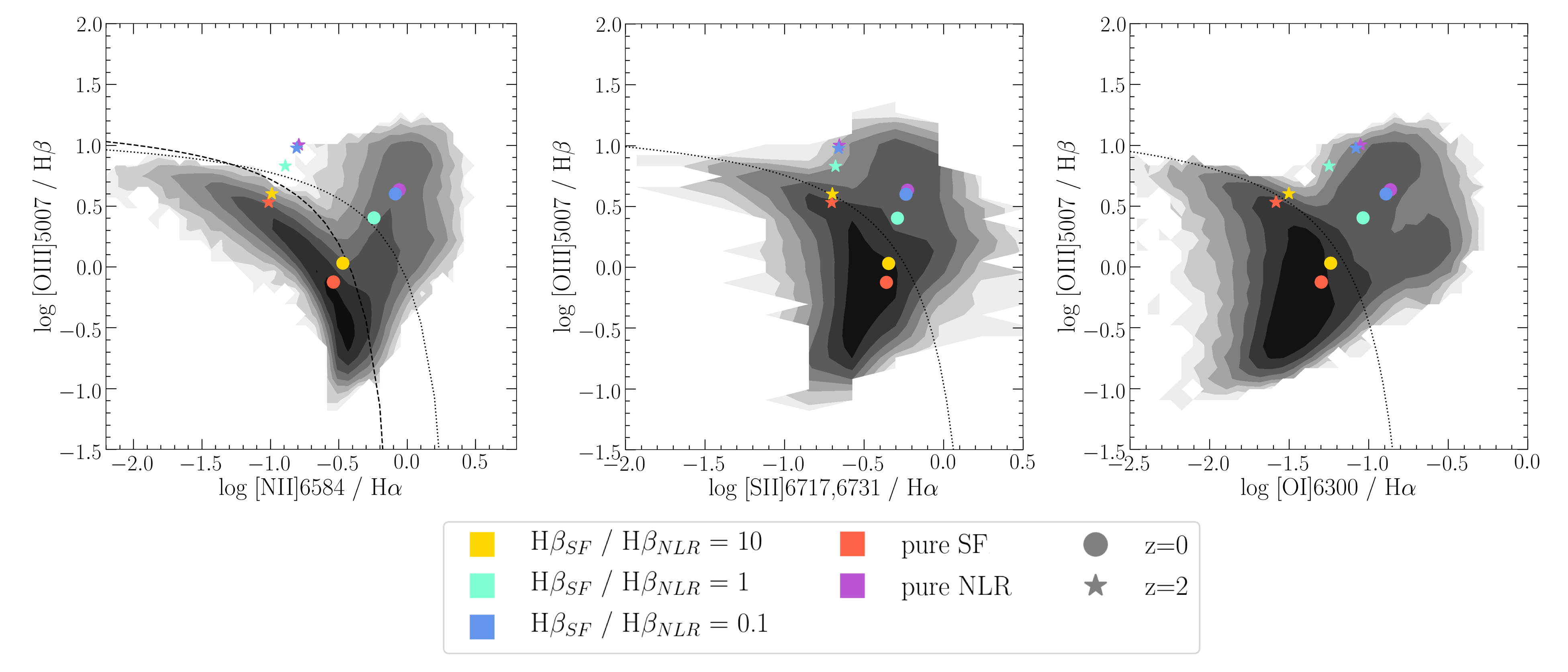}
  \caption{The \NII, \SII\ and \OI\ BPT diagrams showing the coverage of galaxies within SDSS as grey filled contours.  Violet symbols show pure NLR models chosen to represent typical AGNs at $z=0$ (circles) and $z=2$ (stars), while orange symbols show the models chosen to represent typical star-forming galaxies at the same redshifts.  The parameters chosen to represent these galaxies are motivated by the IllustrisTNG-100 simulation (see text for details).  The points spanning between the two extremes show different contributions of NLR model to the corresponding star-forming model, parametrized by the fractional contribution of NLR to the \Hbeta\ flux (as described in the legend).}
  \label{fig:BPT_grid}
\end{figure*}

To investigate how well we can identify type-2 AGNs and constrain their NLR properties at different redshifts, we choose to model two `typical' star-forming galaxies at $z=0$ and $z=2$, to which we add various levels of AGN contribution.  
We select the input parameters of these models by appealing to the IllustrisTNG-100 cosmological, magneto-hydrodynamic simulation of galaxy formation \citep{Springel2018,Pillepich2018,Naiman2018,Nelson2018,Marinacci2018}. See Hirschmann et al. in prep. for the details of how these parameters have been retrieved \citep[see also][]{Hirschmann2017,Hirschmann2019}.  Fig.~\ref{fig:histo_param_input} shows histograms of selected physical properties of galaxies with stellar masses at least $3\times10^9$\Msun\footnote{The stellar masses that we study are significantly higher than the mass-particle resolution of the IllustrisTNG-100.} extracted from this simulation.


We fix the stellar masses of both $z=0$ and $z=2$ galaxies to $\log(\MstarInLog/\Msun) = 10.5$, corresponding to just below the turnover of the stellar mass function of star-forming galaxies measured over this redshift range \citep[e.g.][]{Tomczak2014}. Setting the ages of the oldest stars to 10~Gyr at $z=0$ and 1~Gyr at $z=2$ ensures that the objects are younger than the Universe. The star formation history is parametrized as constant, with an ongoing uncoupled 10-Myr burst of star formation.  We take the current star formation rates of the $z=0$ and 2 galaxies to correspond to the mean values from the Illustris simulation at these redshifts, i.e., $\sfrInLog\approx0.4$ and $1.3\Msun$\,yr$^{-1}$, respectively.  The distributions of \Zhii\ and \logUs\ for star-forming galaxies at $z\sim0$ and 2 in the Illustris simulation are fairly broad and allow for some freedom in choosing typical values.  We adopt values of \Zhii\ and \logUs\ such that the purely star-forming models lie on the sequence of SDSS star-forming galaxies in the classical \NII/\Halpha\ BPT diagram (see Fig.~\ref{fig:BPT_grid}).  For the AGN component, we adopt the mean \ZAGN\ from Illustris at each redshift and tweak \logUsAGN\ to ensure that the models sample the AGN portion of the \OI/\Halpha\ BPT (rather than the LINER/shock region at lower \OII/\Hbeta\ and higher \OI/\Halpha). 

Among the other parameters remaining to be specified, the C/O abundance ratio is fixed to solar, the hydrogen densities in the \HII-region and NLR gas to $\log(\nH/\textrm{cm}^{-3}) = 2$ and 3, respectively, and the upper mass cutoff of the IMF to 100\Msun\ (see Sections~\ref{section:stars_hii} and \ref{section:agn_nlr}). We further fix  the dust-to-metal mass ratio to $\xid=\xidAGN=0.3$, the V-band attenuation optical depth and its fraction arising from the dust in the ambient ISM to $\tauV=0.3$ and $\mu=0.4$ respectively. Finally, we set the power-law slope of the accretion-disc emission model to $\PLalpha=-1.7$, in the middle of the range probed by the AGN-NLR model (Section~\ref{section:agn_nlr}). Table~\ref{tab:grid_table} summarises the parameters of these idealised models. 

When combining the \HII-region and NLR models, we consider three types of galaxies at each redshift: one for which line emission is dominated by star formation, one for which it is dominated by the NLR, and one with equal contribution.  We set
the ratio of \Hbeta\ luminosities from \HII\ regions and the NLR be $\Hbeta^\txn{HII}/\Hbeta^\txn{NLR}= 0.1$, 1 and 10.  This ensures that our models bridge the space between the regions dominated by star-forming galaxies and AGNs in the BPT. 
We ensure the highest value of \Lacc\ (obtained for $\Hbeta^\txn{HII}/\Hbeta^\txn{NLR}= 0.1$) is lower than the Eddington luminosity,
\begin{equation}\label{eq:ledd}
L_{\mathrm{Edd}} = \dfrac{4 \pi G c m_\mathrm{p}}{\sigma_\mathrm{e}} \Mbh \approx 1.26 \times 10^{38} (\Mbh/\Msun)\,\mathrm{erg\,s}^{-1}\,,
\end{equation}
where \Mbh\ is the black-hole mass, $c$ the speed of light, $m_\mathrm{p}$ the proton mass and $\sigma_\mathrm{e}$ the Thomson-scattering cross-section for electrons. In the local Universe, the mass of the central black hole in active galaxies tends to scale with total stellar mass as $\Mbh\sim0.025\Mstar$ \citep[e.g.][]{Reines2015}. Hence, the condition $\Lacc<L_\txn{Edd}$ requires the galaxy stellar mass to be at least $\sim40$ times the corresponding black-hole mass in equation~(\ref{eq:ledd}), which is the case in our models at both redshifts. 

Finally, we also add flat noise to the synthetic spectra, which we parametrize in terms of the observed \Hbeta\ signal-to-noise ratio, S/N(\Hbeta). The standard deviation of the noise per pixel, $\sigma_\mathrm{N}$, is linked to S/N(\Hbeta) through the formula \citep[e.g.][]{Hagen2007}
\begin{equation}\label{eq:noise}
\sigma_\mathrm{N} = \sqrt{\dfrac{2 w_\beta \delta \sqrt{\pi} A_\beta^{2}}{3 \mathrm{[S/N(\Hbeta)]}^{2}}}\,,
\end{equation}
where $w_\beta$ is the \Hbeta\ line width, $\delta$ the pixel width and $A_\beta$ the amplitude of the line. Throughout this work, we investigate the effect of S/N(\Hbeta) on derived properties.
        
At low $\textrm{S/N(H}\beta)$, the perturbed line fluxes for weak lines can fluctuate significantly. To mitigate this effect, we produce 10 spectra for each $\textrm{S/N(H}\beta)$ threshold and average the results after having fitted to the spectra with \beagle.


\subsection{Fitting with \beagle}\label{section:fitting}

Our aims when fitting with \beagle\ are two-fold.  First we hope to distinguish objects with an obscured-AGN component contributing to the emission-line fluxes.
Additionally we wish to constrain the gas properties of the NLR and star-forming regions.  Our base set of observables are the lines used to make the three emission-line diagnostic diagrams identified by \cite{BPT} for distinguishing the dominant ionizing source, namely \OIII, \OI, \Halpha, \Hbeta, \NII\ and \SII.  We add the \OII\ emission line doublet to gain constraints on the ionization state of the gas (most directly together with \OIII\ for constraints on \logUs\ and \logUsAGN). 
We measure these line fluxes and associated uncertainties from the noisy spectra described in section~\ref{section:grid_description} using \pyspeclines.\footnote{\href{https://github.com/jacopo-chevallard/PySpecLines}{www.github.com/jacopo-chevallard/PySpecLines}} Specifically, we fit Gaussian profiles to each of the lines, simultaneously fitting multiple lines that are close or overlapping in the spectrum. We employ the \textsc{mcmc} option in \pyspeclines\ for the line fitting to ensure realistic uncertainty estimates. 

From a first analysis we found that fitting to individual line fluxes produces biased estimates of AGN parameters, and so we construct the likelihood using line ratios particularly sensitive to the estimated physical parameters.  In particular, we use: the BPT line ratios \OI/\Halpha, \NII/\Halpha, \SII/\Halpha\ and \OIII/\Hbeta; \Hbeta/\Halpha\ to provide constraints on dust attenuation; \OI/\OII\ and \OII/\OIII\ to provide constraints on the ionization parameters.  To constrain the absolute normalisation of the line fluxes, we also include the comparison of  \Halpha\ flux in the likelihood calculations. The above line ratios produce improved estimates of various parameters thanks to the higher weighting of lines in the partially ionized zone (\NII, \SII, \OI) to the likelihood evaluation.  When these lines are faint and not fitted as line-ratios but rather individually, the information in the likelihood is dominated by stronger lines which contain less information about the NLR properties (\Halpha, \Hbeta, \OII). We make the simplifying assumption that the line ratios are independent. The method of using line ratios in this way is similar to that adopted by other works, such as \citet{Perez-Montero2019}.  

Within the likelihood determination, we assume that the measurement of the ratio is Gaussian-distributed, with $\mu$ the mean and $\sigma$ the standard deviation of the distribution.  The ratio between two Gaussian-distributed variables, $Z=X/Y$ [where $X\sim\mathcal{N}(\mu_x,\sigma^2_x)$ and $Y\sim\mathcal{N}(\mu_y,\sigma^2_y)$],  is strictly described by a Cauchy distribution when the mean of each distribution is zero.  However, the distribution of the ratio of two independent normal random variables can be approximated as a Gaussian with variance:
\begin{equation}
    \sigma^2_z = \frac{\mu_x^2}{\mu_y^2}\left(\frac{\sigma_x^2}{\mu_x^2} + \frac{\sigma_y^2}{\mu^2_y}\right)
\end{equation}
when $0<\sigma_x/\mu_x<\lambda\leq1$ and $0<\sigma_y/\mu_y\leq\sqrt{\lambda^2-\sigma_x^2/\mu_x^2}<\lambda$, where $\lambda$ is a constant between zero and one \citep{Diaz-Frances2013}, and $\sigma/\mu$ is equal to the inverse of the S/N on the given parameter.  We always choose ratios with the lower S/N line as the numerator, which safely respects the condition on $\sigma_x/\mu_x$.  Re-writing the central part of the second condition using $\sigma_y/\mu_y=\frac{1}{N}\sigma_x/\mu_x$ (which is equivalent to saying the signal-to-noise on $y$ is a multiple of $N$ times higher than the signal-to-noise on $x$), we have $\sqrt{(N^2+1)}\sigma_x/\mu_x<N\lambda$, which is trivially true if $N>1$ and $\sigma_x/\mu_x<\lambda$.
Therefore, we are generally within the regime where we can safely model the distribution of the line ratios as a Gaussian. In practice, the lines on the numerator do not always satisfy $\sigma_x/\mu_x<1$. However, when they have very low S/N (high $\sigma_x/\mu_x$), their weighting within the likelihood is much lower and so this should not significantly affect the results.



\begin{table*}
\centering
\caption{Uniform prior limits and fixed values applied to the three different fits for different parameters in \beagle. }
\begin{tabular}{ C{0.45\columnwidth} C{0.45\columnwidth} C{0.45\columnwidth}C{0.45\columnwidth}} 
\toprule

 \multicolumn{1}{c}{Parameter}	   & \multicolumn{1}{c}{Fit \PLalpha}  & \multicolumn{1}{c}{Fit \xidAGN} & \multicolumn{1}{c}{Fixed \PLalpha and \xidAGN}\\
 
\midrule

$\Mtot=\log(\MtotInLog/\Msun)$ & Uniform $\in [7,13]$ & $\in [7,13]$ & $\in [7,13]$\\

$\sfr=\log(\sfrInLog/\Msun \yr^{-1})$ & Uniform $\in [-4.,4.]$ & Uniform $\in [-4.,4.]$ & Uniform $\in [-4.,4.]$\\

$\log(\Zhii/\Zsun)$ & Uniform $\in [-2.2,0.24]$ & Uniform $\in [-2.2,0.24]$ & Uniform $\in [-2.2,0.24]$\\

$\log(\t/\yr)$ & Uniform $\in [6,11]$ & Uniform $\in [6,11]$ & Uniform $\in [6,11]$\\

\logUs & Uniform $\in [-4,-1]$ & Uniform $\in [-4,-1]$ & Uniform $\in [-4,-1]$\\

\xid & Fixed to 0.3  & Fixed to 0.3  & Fixed to 0.3 \\

\tauV & Uniform $\in [0, 2]$  & Uniform $\in [0, 2]$  & Uniform $\in [0, 2]$ \\

\mud & Uniform $\in [0,1]$ & Uniform $\in [0,1]$ & Uniform $\in [0,1]$\\

$\log (\Lacc\/\Lsun)$ & Uniform $\in [40,48]$ & Uniform $\in [40,48]$ & Uniform $\in [40,48]$\\

\xidAGN\ & Fixed to 0.3 & Uniform $\in [0.1,0.5]$ & Fixed to 0.3 \\

\logUsAGN & Uniform $\in [-4,-1.5]$ & Uniform $\in [-4,-1.5]$ & Uniform $\in [-4,-1.5]$\\

$\log(\ZAGN/\Zsun)$& Uniform $\in [-2.2,0.24]$ & Uniform $\in [-2.2,0.24]$ & Uniform $\in [-2.2,0.24]$\\

\PLalpha\ & Uniform $\in [-2.0,-1.2]$  & Fixed to -1.7 & Fixed to -1.7 \\

\bottomrule
\end{tabular}
\label{tab:test_scenario_beagle_parameters}
\end{table*}




\subsection{High-quality spectra with S/N(H$\beta$)=100}\label{section:highqual}


\subsubsection{AGN versus star-formation dominated spectra}\label{sec:agnsfdom}

\begin{figure*}
  \centering
  \subfigure[$z=0$]{\includegraphics[width=3.4in,trim={1cm 0cm 4cm 5cm},clip]{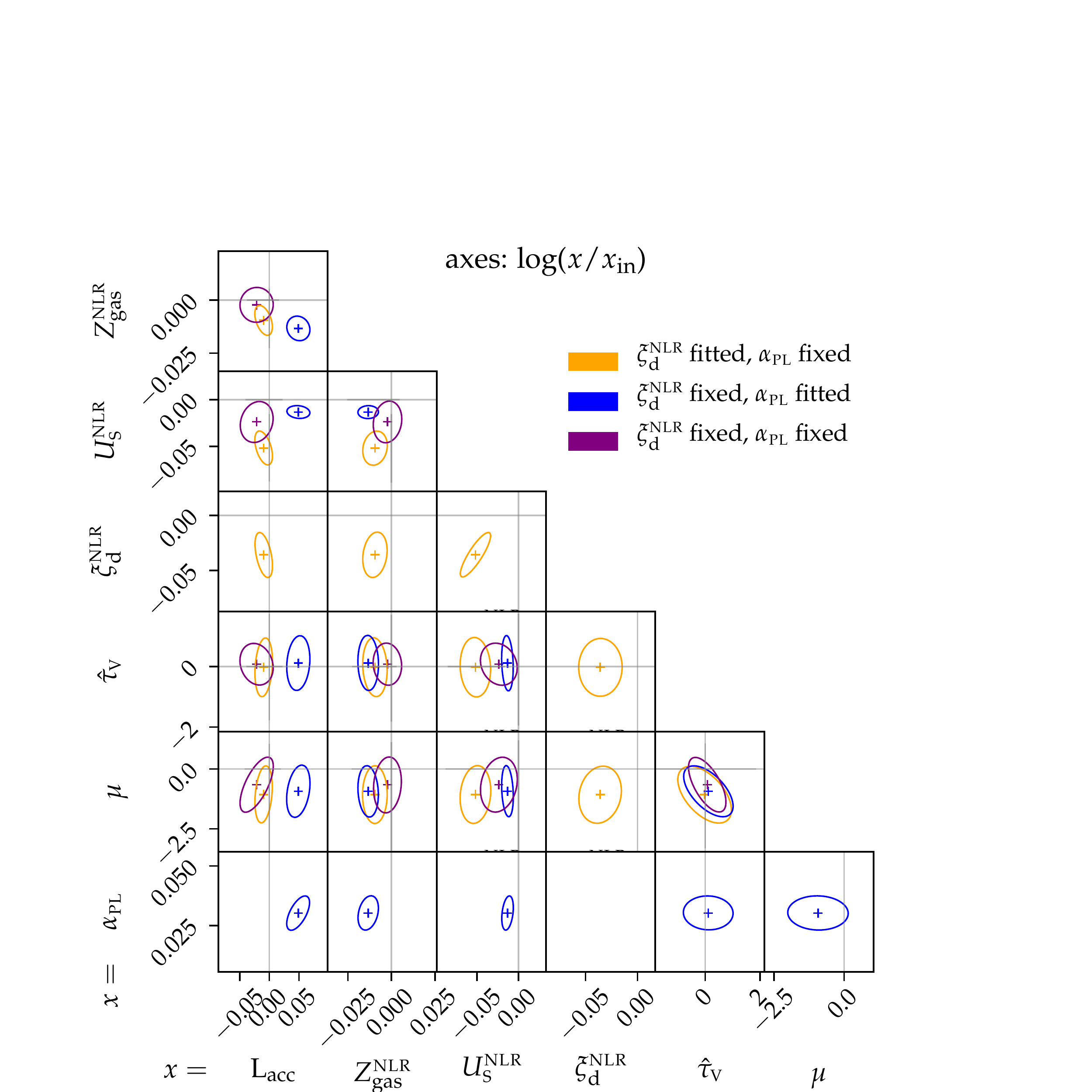}}
  \subfigure[$z=2$]{\includegraphics[width=3.4in,trim={1cm 0cm 4cm 5cm},clip]{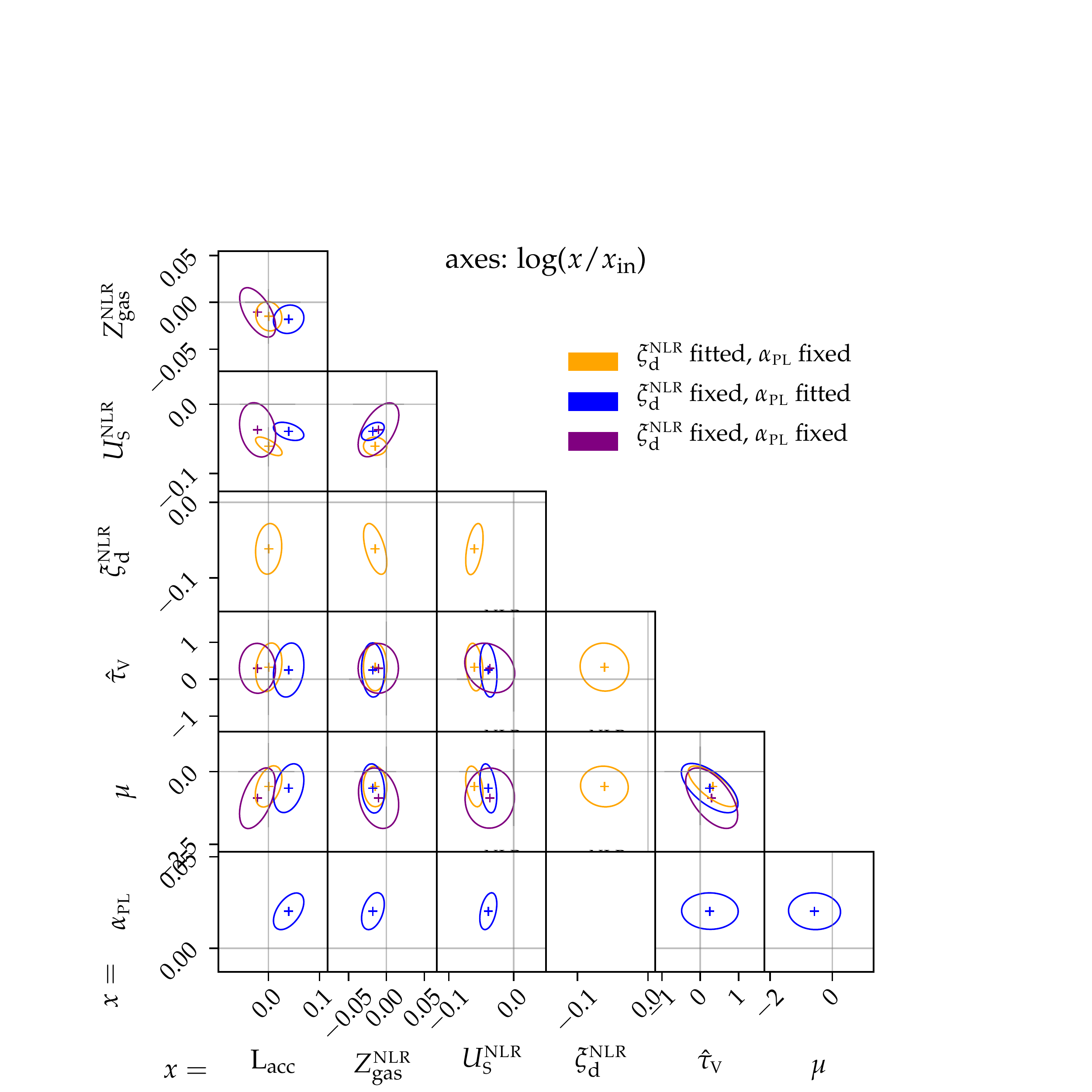}}
  \caption{Triangle plots of the parameter retrieval for the $z=0$ (left) and $z=2$ (right) NLR-dominated galaxies with S/N$\sim100$ in \Hbeta.  Only NLR parameters are displayed even though SF parameters are also included in the fits.  The triangle plots display the average parameter constraints relative to the input values from fits to 10 realisations of noisy spectrum.  The crosses show the average bias for each parameter pair, while the ovals show the $1\sigma$ contour of the bi-variate Gaussian fitted to the joint posteriors (see text for details about how this plot is made).  These show an approximate representation of the biases and degeneracies between different parameters.  We display the results for different combinations of fixed or fitted \xidAGN\ and \PLalpha, as specified in the legend.}
  \label{fig:average_triangle_AGN}
\end{figure*}

\begin{figure*}
  \centering
  \subfigure[$z=0$]{\includegraphics[width=3.4in,trim={1cm 0cm 4cm 5cm},clip]{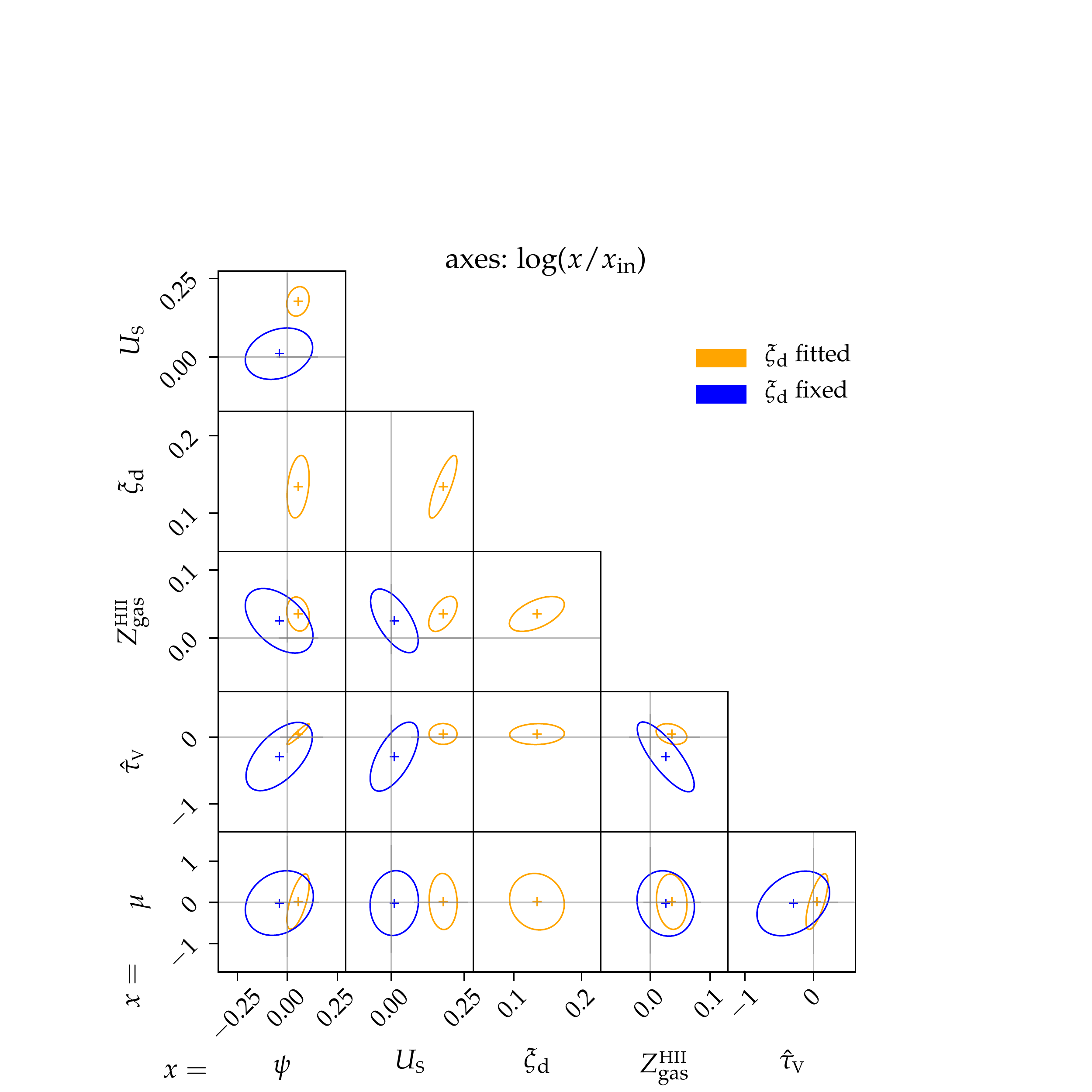}}
  \subfigure[$z=2$]{\includegraphics[width=3.4in,trim={1cm 0cm 4cm 5cm},clip]{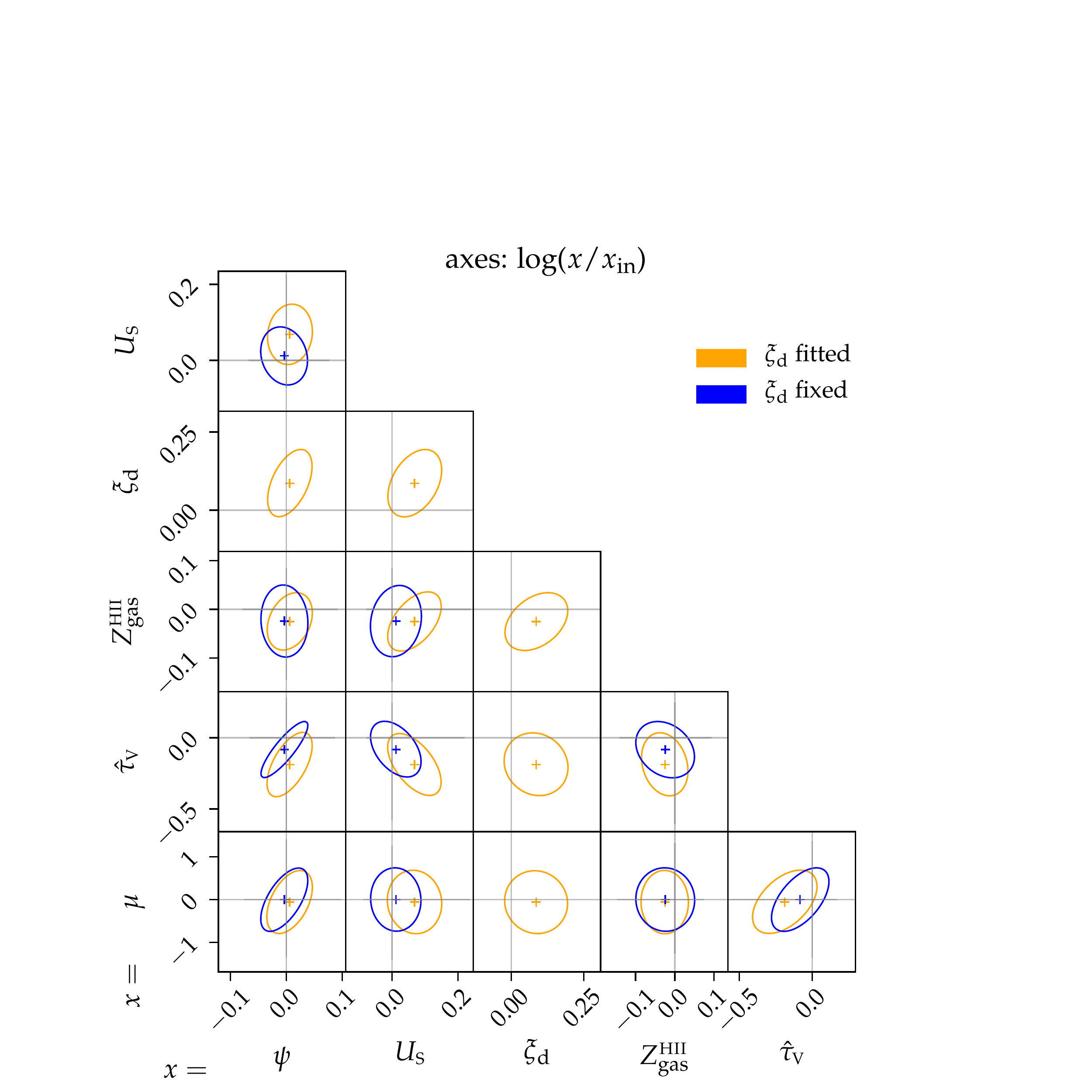}}
  \caption{As in fig.~\ref{fig:average_triangle_AGN}, but showing triangle plots of the parameter retrieval for the $z=0$ (left) and $z=2$ (right) star-formation dominated galaxies.  We display only SF parameters, though NLR parameters were also included in the fits.  Results when \xid\ is both fixed and fitted are displayed according to the legend. }
  \label{fig:average_triangle_SF}
\end{figure*}

We start by investigating the idealised scenario of high-S/N spectra with full coverage of the lines of interest.  From our grid of models, we test the retrieval of physical parameters for the AGN and star-formation dominated cases ($\Hbeta_\textsc{sf}/\Hbeta_\textsc{nlr} = 0.1$ and 0.9, respectively) with high S/N spectra [S/N(H$\beta$)=100].  At this S/N in \Hbeta\, we reach $\textrm{S/N}>10$ in all lines used in the analysis. We fit all components (\HII, stellar continuum and NLR) to each spectrum to ensure that the parameters of the dominant component are not biased by the expanded parameter space including the  rest of the components.

Fig.~\ref{fig:average_triangle_AGN} summarises the joint posterior probability distributions derived from \beagle\ fits to the $z=0$ and $z=2$ AGN-dominated spectra.  We display only parameters that affect the NLR emission in the model, even though \HII\ and stellar parameters were fitted.  The plot is an effective average of the posteriors derived for each of the 10 realisations at each grid point (where we take the same input spectrum and produce 10 realisations with random noise).  For each noisy spectrum, we take samples from the posterior probability (derived using \beagle) for each fitted parameter before computing the logarithm of the ratio to the corresponding input value.  We then fit a bi-variate Gaussian to the joint distribution of logarithmic ratios for each parameter pair.  The mean of the 10 bi-variate Gaussian centres are plotted as crosses. 
The ovals show the average of the bi-variate Gaussian distributions of the fits to the 10 noisy spectra (specifically, the covariance matrices were constructed using the mean of the corresponding entries in the individual-fit covariance matrices).

The results are displayed for three different parameter configurations; varying \PLalpha\ while fixing \xidAGN (yellow), varying \xidAGN\ while fixing \PLalpha (blue), and fixing both of these parameters (red).  We find that varying \xidAGN\ leads to \logUsAGN\ and \ZAGN\ being under-estimated for both the $z=0$ and $z=2$ objects.  \xidAGN\ is clearly poorly constrained and there are degeneracies between \xidAGN\ and \logUsAGN.  The depletion onto dust grains can, in principle, be constrained using observables from elements with different refractory properties.  For example, Oxygen is depleted onto dust grains, but Nitrogen is not.  A standard line ratio used for determining gas-phase N/O ratio is \NII/\OII, because both ions have similar ionization energies. 
We tried explicitly including this ratio in the fits with little effect on the results.  This is because, in the presence of ionizing radiation from an accretion disc, the \NII/\OII\ ratio is heavily affected by the hardness of the incident radiation, and the residual dependence on gas-phase N/O abundance is insufficient to constrain \xidAGN. 

Allowing \PLalpha\ to vary leads to over-estimated \Lacc\ while still under-estimating \ZAGN. 
There are no clear degeneracies that would lead to such a biased \ZAGN\ estimate.  There is a clear degeneracy between \PLalpha\ and \Lacc, resulting from the definition of \Lacc, the integration of the thermal accretion disc model (Fig.~\ref{fig:accretion disc}).  Increasing \PLalpha\ while keeping everything else constant will increase \Lacc.      We find the best recovery of \Lacc, \ZAGN\ and \logUsAGN\ when fixing both \xidAGN\ and \PLalpha.  \cite{Feltre2016} show that increasing \PLalpha\ pushes the models to higher values of \OIII/\Hbeta\ in the \NII/\Halpha\ BPT diagram (see their figure 2), so it may be 
appropriate to vary \PLalpha\ for objects with more extreme measured line ratios. See paper II for a further analysis of how the variation of \PLalpha\ affects line ratios in the models and the fitting to real data.

The fraction of the V-band attenuation optical depth arising from the dust in the ambient ISM, $\mu$, is not well constrained by this set of observables for the $z=0$ galaxy nor for the $z=2$ one.

Fig.~\ref{fig:average_triangle_SF} displays the stellar and \HII\ region parameters for the star-formation dominated $z=0$ and $z=2$ galaxies.  We show results when nebular \xid\ is fixed 
and fitted. We find that \xid\ is poorly constrained, leading to over-estimated \logUs\ and \Zhii, as well as tighter constraints on the biased estimates.  The results are somewhat less biased for the $z=2$ galaxy. 
Finally, we note that for the star-forming dominated case, both in the $z=0$ and the $z=2$ cases, $\mu$ is well constrained.

\subsubsection{Equal contributions by AGN and star formation}

\begin{figure*}
  \centering
  \includegraphics[width=7in,trim={2cm 0cm 3cm 4cm},clip]{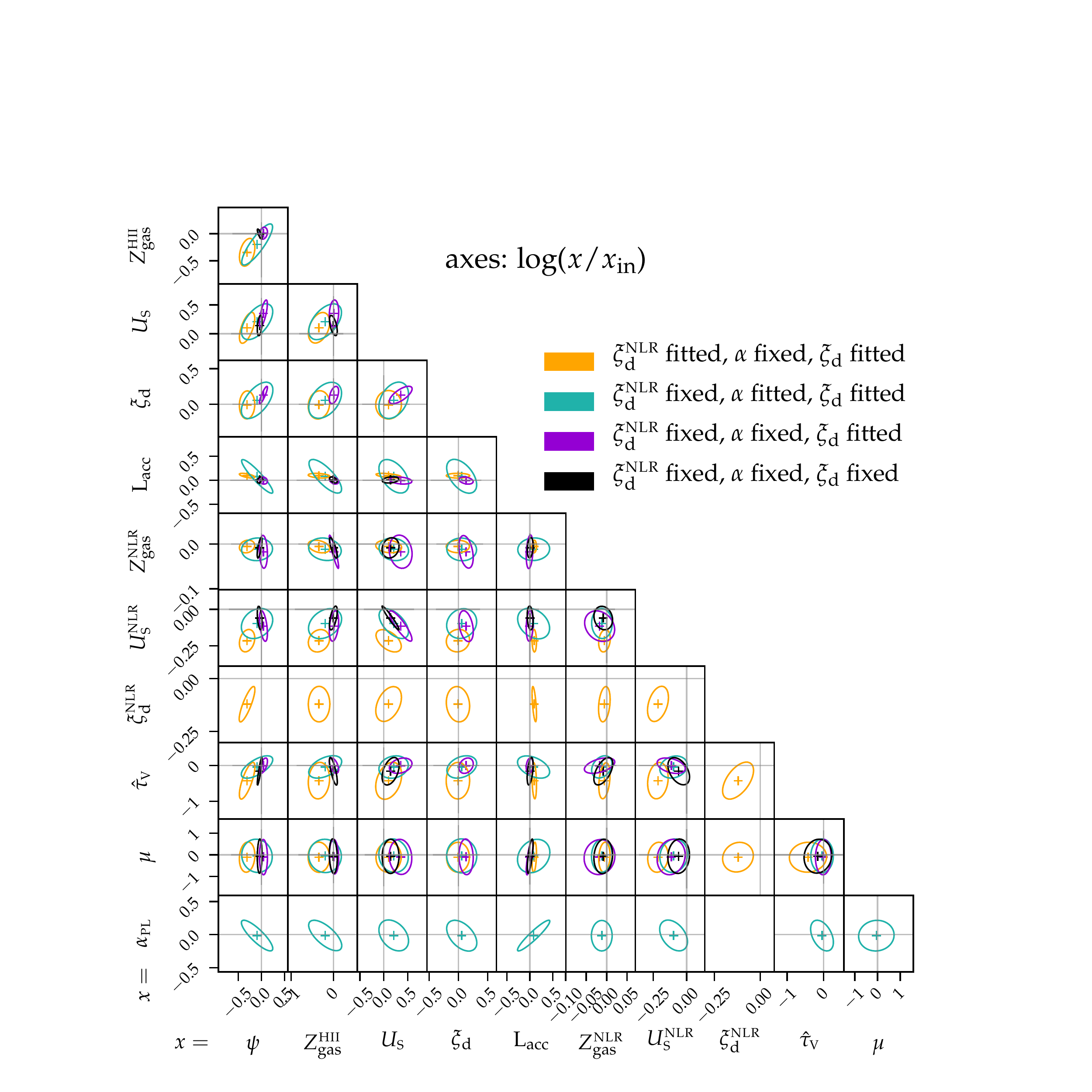}
  \caption{As in fig.~\ref{fig:average_triangle_AGN} but now showing the triangle plots of the parameter retrieval for the $z=0$ galaxy with equal contribution from SF and NLR to \Hbeta\ flux.  We display the results for all fitted parameters, with a range of different configurations for fixing or fitting to \xidAGN, \PLalpha.}
  \label{fig:average_triangle_all_z0}
\end{figure*}

\begin{figure*}
  \centering
  \includegraphics[width=7in,trim={2cm 0cm 3cm 4cm},clip]{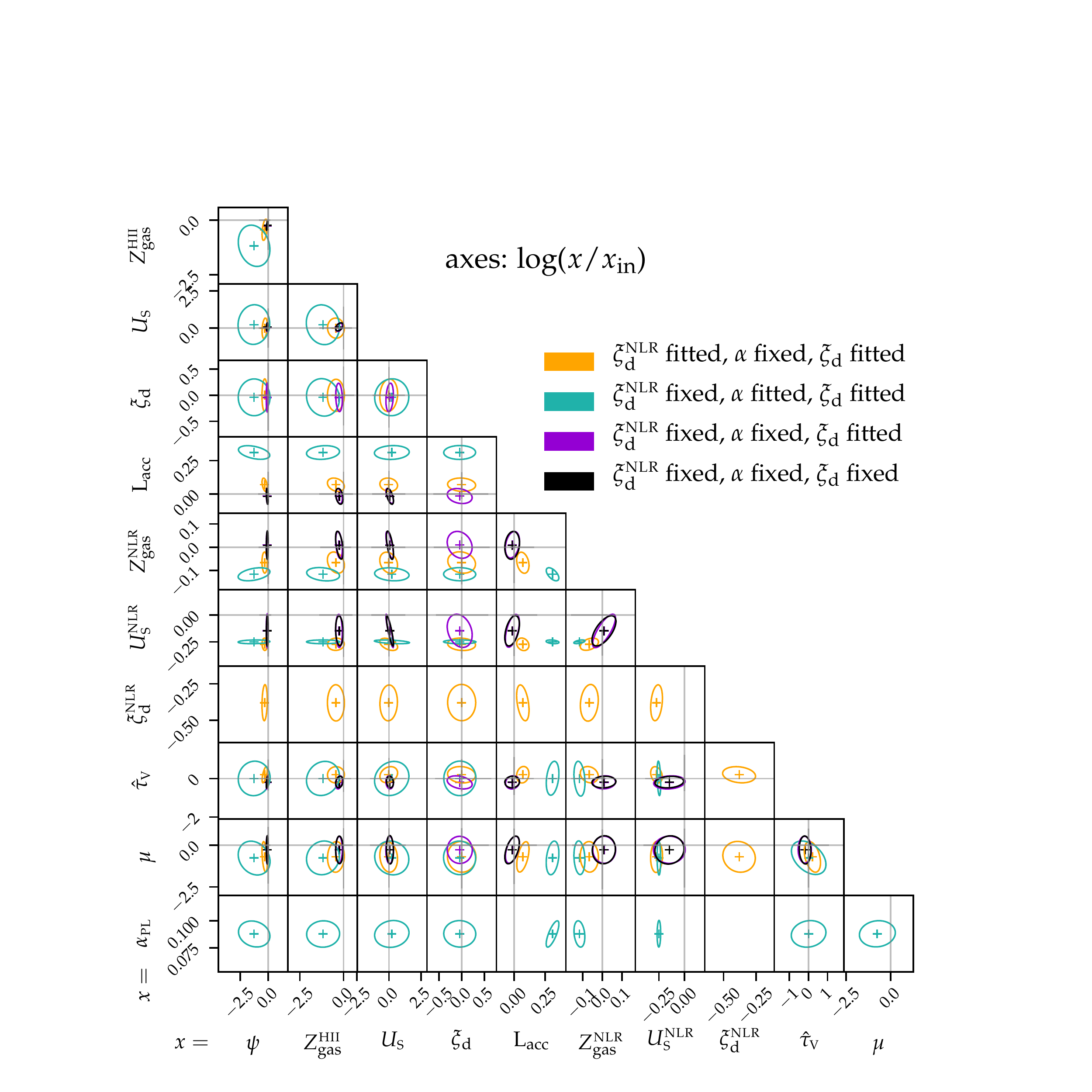}
  \caption{As in fig.~\ref{fig:average_triangle_AGN} but now showing the triangle plots of the parameter retrieval for the $z=2$ galaxy with equal contribution from SF and NLR to \Hbeta\ flux.}
  \label{fig:average_triangle_all_z2}
\end{figure*}

We proceed with the most challenging case of equal contributions of NLR and \HII\ regions to the \Hbeta\ flux.  Figs.~\ref{fig:average_triangle_all_z0} and~\ref{fig:average_triangle_all_z2} show the average joint posteriors for all fitted parameters for the $z=0$ and $z=2$ galaxy, respectively.  We display the constraints for different configurations of fitted parameters, as indicated in the legend.  Notably, the constraints are quite different for the two different objects.  For instance, \PLalpha\ is well recovered for the $z=0$ object, but is significantly biased for the $z=2$ object.  The biased estimates of \PLalpha\ for the $z=2$ object consequently lead to significantly biased estimates of \logUsAGN, \ZAGN\ and \Lacc.  We therefore learn that the accuracy of the \PLalpha\ estimates depends heavily on the region of the observable parameter space that is being probed.  In other words, \PLalpha\ can be well retrieved for objects in certain regions of the \NII\ BPT diagram, probably dependent on the sampling of the parameter space\footnote{An irregular sampling of the parameter space is a natural outcome from physical motivated model grids.} by the models and the degeneracies present in that particular region.  We therefore suggest that \PLalpha\ should be fixed as standard when using this set of observables, unless 
reaching an acceptable fit requires this parameter to be varied.  

A parameter which was found to be problematic in the AGN-dominated fits is \xidAGN (Section~\ref{sec:agnsfdom}), which again is poorly constrained in case of equal contributions by NLR and \HII\ regions. \xidAGN\ is degenerate with \sfr, \Lacc\ and \logUsAGN, meaning that when \xidAGN\ is un-constrained, the recovery of these parameters will also be biased.  Fixing \xidAGN\ somewhat improves the retrieval of the NLR parameters, but for the $z=0$ galaxy, \xid\ is still poorly constrained.  This parameter is degenerate with \logUs\, which is, in turn, degenerate with \logUsAGN. Therefore fixing both \xidAGN\ and \xid\ provide less biased estimates of the ionization parameters of the NLR and \HII\ region emission. We therefore obtain best constraints for both objects when \xid\ and \xidAGN\ are both fixed.

Finally, $\mu$ is well constrained for the $z=0$ object for the different configurations of the fitted and fixed parameters but it is not so well constrained for the same configurations for the $z=2$ object. In order not to be biased by the incertitude on the retrieval of this parameter, we fix the value of $\mu$ in the following sections.


\subsection{Influence of S/N on parameter retrieval}\label{section:influenceSN}

\begin{figure}
  \centering
  \includegraphics[width=3.4in,trim={2.5cm 0cm 3cm 5cm},clip]{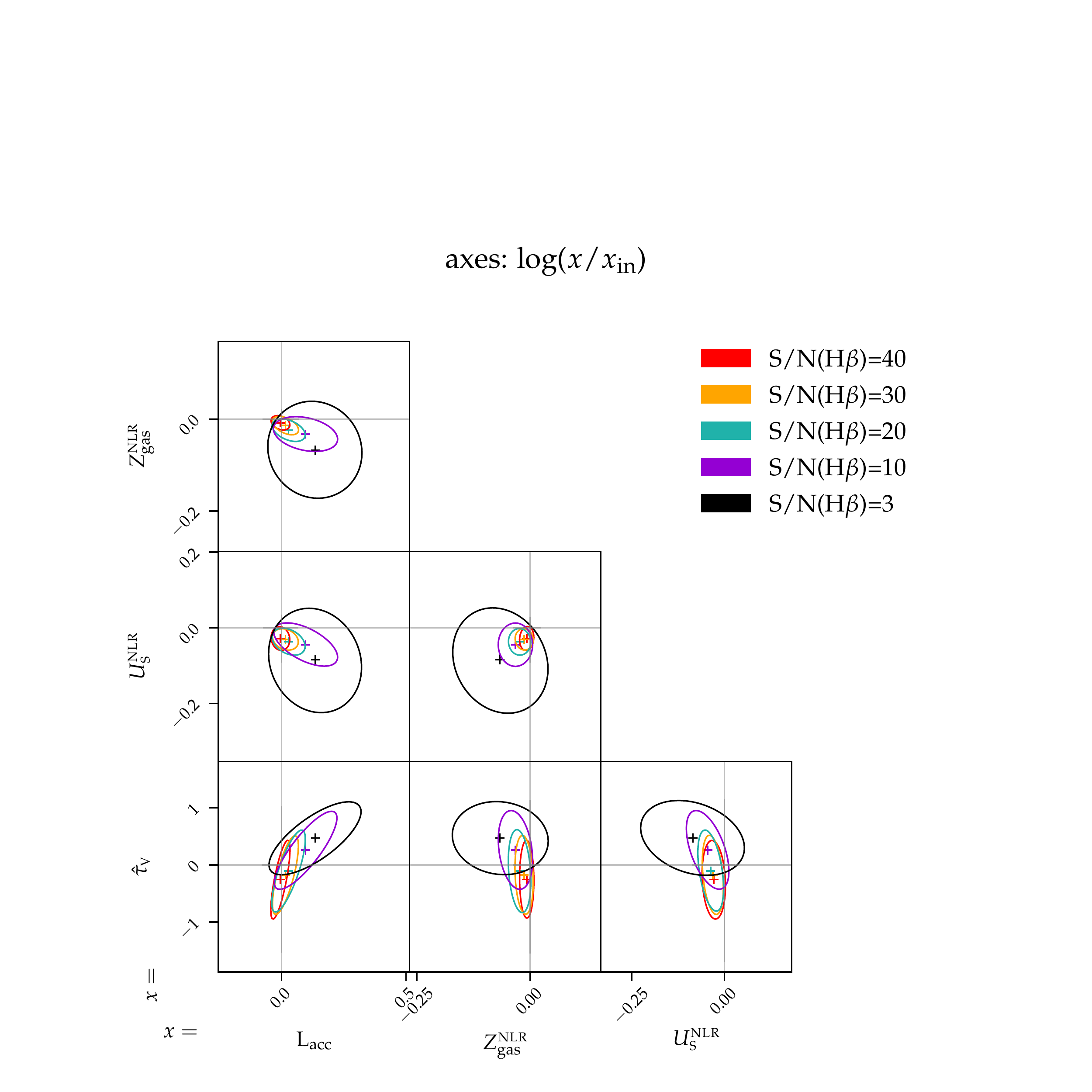}
  \caption{Triangle plots of the NLR parameter retrieval for the $z=0$ AGN-dominated objects. Different colors represent different S/N in emission lines. For a given parameter, the axes show the logarithmic fraction of the output by the input value.}
  \label{fig:average_triangle_AGN_SN}
\end{figure}

\begin{figure}
  \centering
  \includegraphics[width=3.4in,trim={3cm 0cm 4cm 5cm},clip]{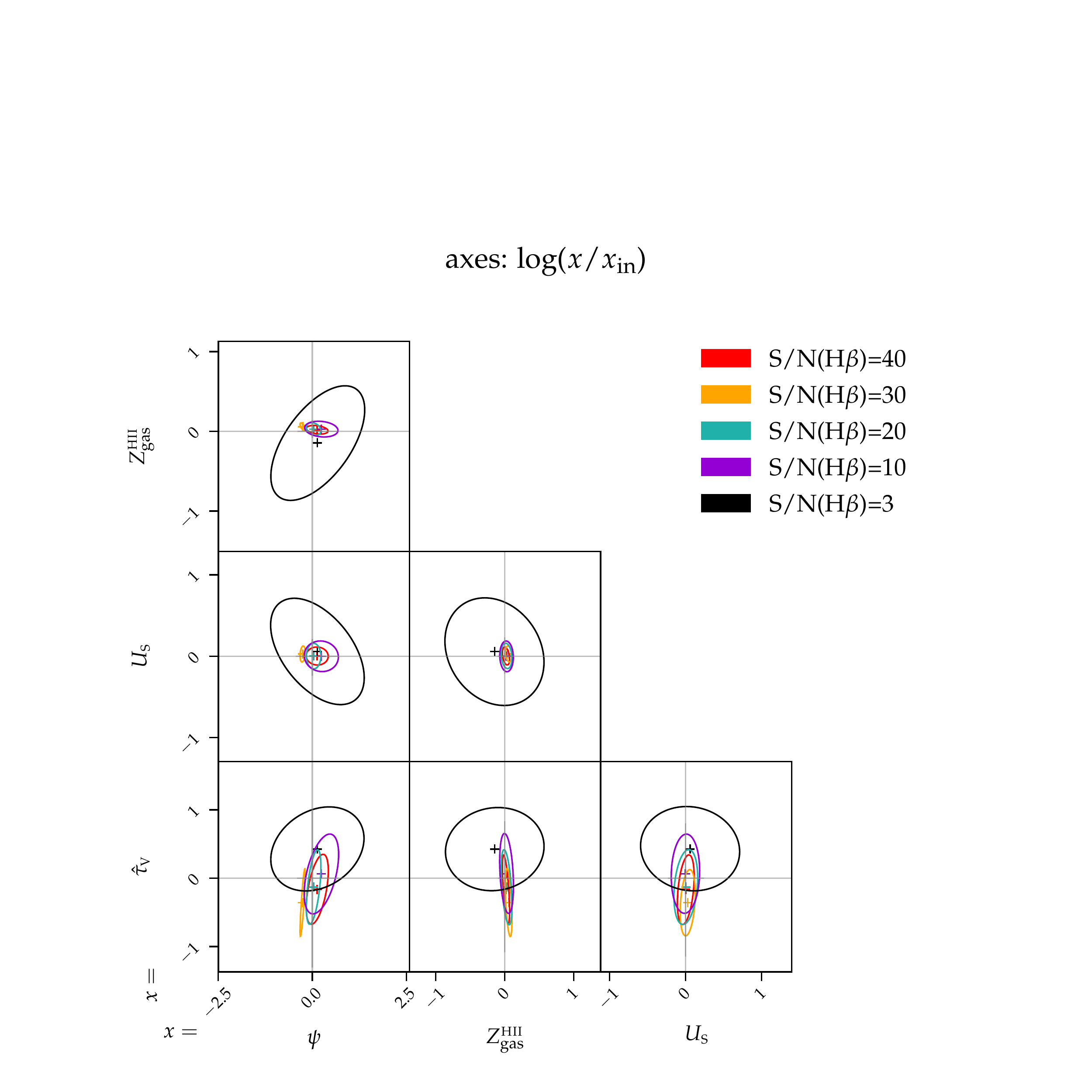}
  \caption{Same as Fig.~\ref{fig:average_triangle_AGN_SN} but for the parameter retrieval of the \HII\ region for SF-dominated objects at $z=0$. }
  \label{fig:average_triangle_SF_SN}
\end{figure}

\begin{figure}
  \centering
  \includegraphics[width=3.4in,trim={3cm 0cm 4cm 5cm},clip]{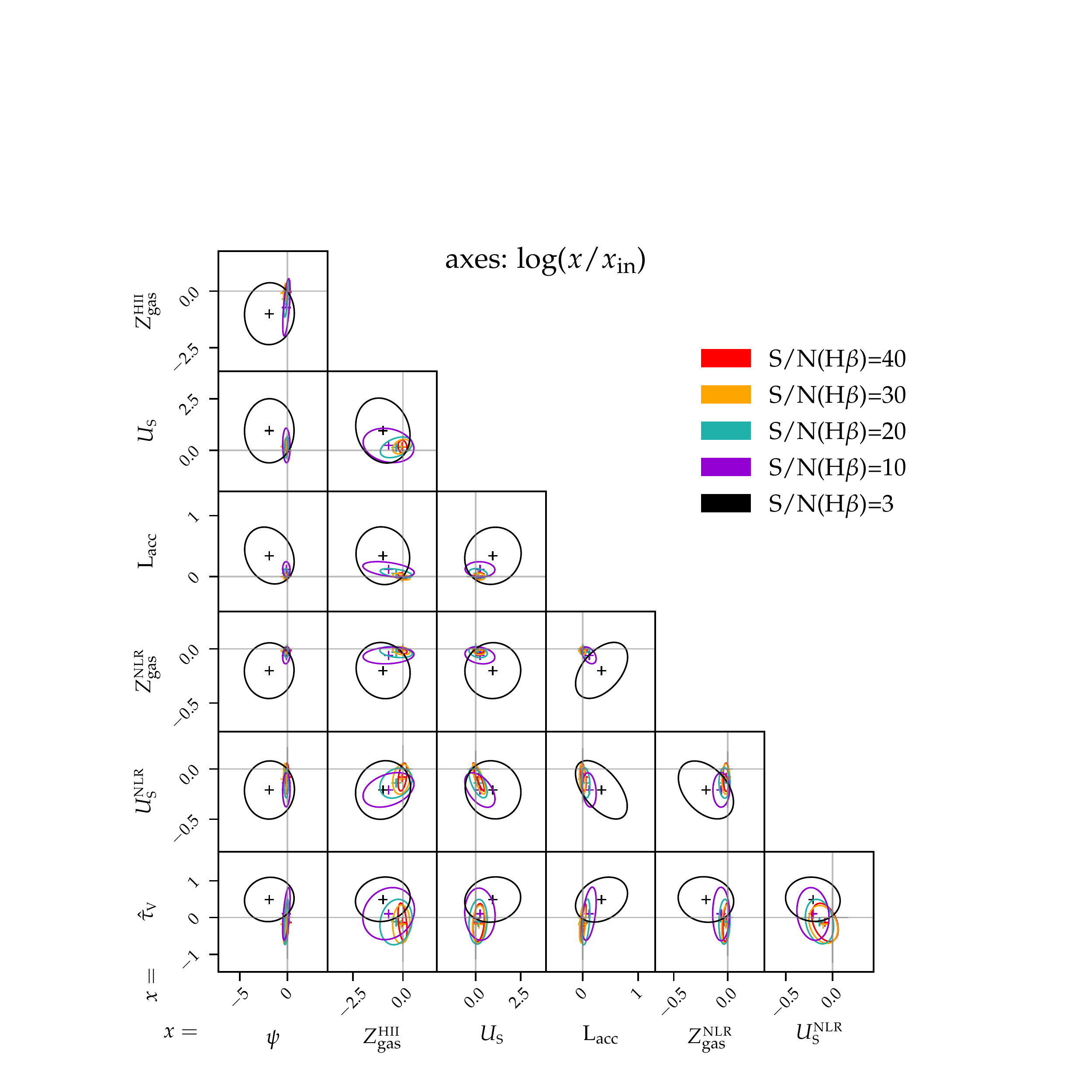}
  \caption{Same as Fig.~\ref{fig:average_triangle_AGN_SN} but for the parameter retrieval of the \HII\ region and the NLR for  objects at $z=0$ with an equal contribution of both regions to \Hbeta.}
  \label{fig:average_triangle_5050_SN}
\end{figure}

We investigate the S/N in emission lines required to derive un-biased parameter estimates.  Based on the results for the $\textrm{S/N(H}\beta)\sim100$ fits, we fix \xid, \xidAGN, \PLalpha\ and \mud. For the AGN-dominated galaxies, the $z=0$ object constraints degrade with S/N more noticeably than for the $z=2$ object, so to be conservative, we only show the $z=0$ results in Fig.~\ref{fig:average_triangle_AGN_SN}.  These results suggest that the AGN parameters are well-constrained at $\textrm{S/N(H}\beta)\gtrsim20$.  With decreasing S/N, degeneracies between \Lacc\ and \tauV, as well as \Lacc\ and \logUsAGN\ become more problematic, causing increasingly biased estimates of the NLR properties.  The specific behaviour with low S/N will be different for each object as our results in Section~\ref{section:retrieval} show how dependent parameter retrieval is on the point in parameter space occupied by the object, because of the very non-linear and irregular coverage of the models in observable space.

The \HII\ region properties for the star-formation dominated objects are well constrained and un-biased down to $\textrm{S/N(H}\beta)\sim10$, as displayed in Fig.~\ref{fig:average_triangle_SF_SN}. 
Disentangling the contributions from star-forming and NLR components requires higher S/N, however, as illustrated by Fig.~\ref{fig:average_triangle_5050_SN}, which shows the parameter retrieval for the case with equal contributions by SF and NLR to \Hbeta.  Parameter retrieval starts to become biased and poorly constrained for $\textrm{S/N(H}\beta)\lesssim30$. 

\subsection{NLR contribution to the spectrum}\label{section:NLRcontribution}

\begin{figure}
  \centering
  \includegraphics[width=3.5in]{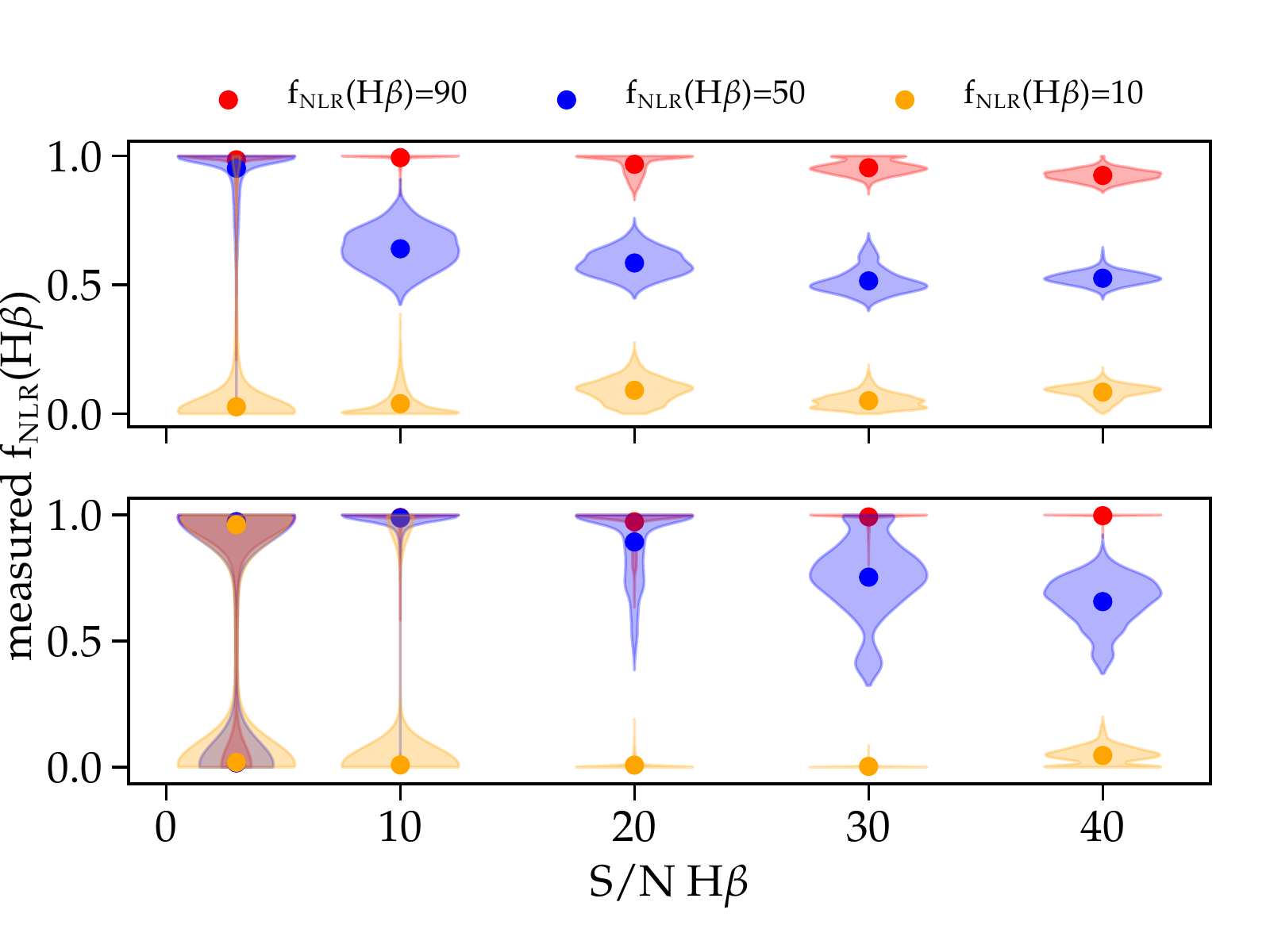}
  \caption{Retrieved fractional contribution of the NLR to the total \Hbeta\ flux for each object in our simulated grid.  The upper (lower) panel shows the retrieval for the typical $z=0$ ($z=2$) objects with different input fractional NLR contribution to \Hbeta\ as displayed in the legend.}
  \label{fig:SN_frac_Hb}
\end{figure}

\beagle\ can be used to identify 
the potential contribution by an AGN to the emission-line luminosities of a galaxy.
To test how well \beagle\ can 
detect the presence of an AGN, we show the retrieved fractional NLR contribution to \Hbeta\ flux as a function of the S/N on \Hbeta\ in Fig.~\ref{fig:SN_frac_Hb}.  If we wish only to identify objects with a significant contribution from the NLR to \Hbeta,  a S/N of 3 in \Hbeta\ is sufficient for our adopted properties of a typical $z=0$ galaxy.  However, this S/N is insufficient to reliably estimate the fractional contribution to \Hbeta, which is over-estimated if stars and the AGN contribute in equal amounts to the line flux.  

For our $z=2$ galaxy, the fractional contribution of the NLR to \Hbeta\ is completely unconstrained at S/N$(\Hbeta)=3$. For S/N$(\Hbeta)\sim10$,  the NLR contribution is over-estimated for the mixed case.  However, only a small probability of a high NLR contribution to the star-formation dominated spectrum (seen from the low fractional area of the yellow curve above 0.5) means that even at this reasonably low S/N, a fairly clean sample of objects with significant contribution of NLR to \Hbeta\ flux can selected. Being more conservative, S/N$\sim20-30$ is required to firmly identify objects with possible NLR contribution, as well as constraining that contribution. It is harder to distinguish NLR contribution in the $z=2$ case because the metallicity is lower, and so key line fluxes such as \OIII, \OII\ or \OI\ are also much fainter with respect to \Hbeta.  This leads to much more overlap between the star-forming and NLR models in this region of the BPT diagram (see Fig.~\ref{fig:BPT_grid}), making separation of the NLR and star-forming contributions more challenging.  The addition of other line ratios may improve the ability of finding NLR contribution at low metallicity.  

\section{Discussion}
\label{section:discussion}

Other works have investigated the derivation of NLR physical parameters from emission lines.  Throughout the discussion, we compare the modelling assumptions we make to those of other works, with the aim of outlining when and why different models will provide different (or similar) properties for the same data sets. The works we compare to are: the strong-line calibrations of \cite{Storchi-Bergmann1998} and \cite{Dors2021}, and the Bayesian approaches of \cite{Perez-Montero2019} and \cite{Thomas2018}.  Before making comparisons, we summarise the different approaches in more detail here. 

\subsection{Different approaches to deriving NLR physical properties}\label{section:approachesNLR}

\subsubsection{Storchi-Bergmann et al. 1998}

The first calibration of oxygen abundances based on strong optical line emission from NLRs was provided by \cite{Storchi-Bergmann1998}.  They used \cloudy\ to produce a grid of emission line fluxes for an incident radiation field described by the segmented power-law of \cite{Mathews1987} (which is shown in comparison to that used in the F16 models in Fig.~\ref{fig:accretion disc}).  The grid covers a range of oxygen abundances and ionization parameters.  Although not explicitly stated, 
the ionization parameter is likely that at the inner edge of the cloud (rather than at the radius of the Str\"omgrem sphere as for the F16 model grid).  They account for both primary and secondary nitrogen, i.e. dependent on the oxygen abundance, following $\log(\textrm{N/O}) = 0.96[\logOH]-9.29$.  They also account for depletion of refractory elements from the gas phase onto dust grains following the values of the
observed abundance of the interstellar medium from \citet{CowieSongaila1986}.
The following equations describe the two calibrations for the gas-phase oxygen abundance derived by \cite{Storchi-Bergmann1998}:
\begin{eqnarray}
       \begin{array}{l@{}l@{}l}
\rm (O/H)_{SB98,1} & = &  8.34  + (0.212 \, x) - (0.012 \,  x^{2}) - (0.002 \,  y)  \\  
         & + & (0.007 \, xy) - (0.002  \, x^{2}y) +(6.52 \times 10^{-4} \, y^{2}) \\  
         & + & (2.27 \times 10^{-4} \, xy^{2}) + (8.87 \times 10^{-5} \, x^{2}y^{2}),   \\
     \end{array}
\label{eq:sb1}
\end{eqnarray}

\noindent where $x$ = [N\,{\sc ii}]$\lambda6548,\lambda6584$/H$\alpha$ and 
$y$ = [O\,{\sc iii}]$\lambda4959,\lambda5007$/H$\beta$ and

\begin{eqnarray}
       \begin{array}{lll}
(\rm {O/H})_{{\rm SB98,2}}   & = &  8.643 - (0.275 \, u) + (0.164 \, u^{2})   \\  
          & + & (0.655 \, v) - (0.154 \, u v)  - (0.021 \, u^{2}v) \\  
          & + & (0.288 v^{2}) + (0.162 u v^{2}) + (0.0353 u^{2}v^{2}),   \\
     \end{array}
\label{eq:sb2}
\end{eqnarray}

\noindent where $u$ = log(\OII/[O\,{\sc iii}]$\lambda4959,\lambda5007$)  and $v$ = log([N\,{\sc ii}]$\lambda6548,\lambda6584$/H$\alpha$).  The term (O/H) above corresponds to 12+log(O/H). Both calibrations are valid for $\rm 8.4 \: \leq \: 12+log(O/H) \:  \leq \: 9.4$.

The calibrations are appropriate for $\nH = 300\, \textrm{cm}^{-3}$, but they also provide a correction factor  for different densities:

\begin{equation}
    (\textrm{O/H})_{final} = (\textrm{O/H}) - 0.1\,\log(\nH/300)
\end{equation}

\subsubsection{Dors 2021}\label{sec:Dors21}

The first strong-line calibration based on gas-phase oxygen abundance estimates derived from the $T_\txn{e}$ method (or electron temperature method) for NLRs was presented in \cite{Dors2021}.  From the $T_\txn{e}$-derived gas-phase oxygen abundances of a sample of AGNs\footnote{Selected from the location of the galaxies in the standard BPT diagrams, see \cite{Dors2020} for the details of the target selection.}, 
\cite{Dors2021} provide a 2D surface fitted to the objects within a 3D space defined by \logOH, $R_{23}$ and $P$ (defined below):
\begin{equation}
    R_{23}=\dfrac{\OII+\OIIIall}{H\beta}
    \label{eq:R23}
\end{equation}
\noindent

\begin{equation}
    P = \frac{\OIIIall/\Hbeta}{R_{23}}
    \label{eq:dors21_p}
\end{equation}
\noindent
The $R_{23}$ ratio is often used as an oxygen abundance indicator but is known to be dependent on ionization parameter, as well as the hardness of the ionizing radiation \citep{Pilyugin2000}, while $P$ is sensitive to radiation hardness \citep{Pilyugin2001}.  The resulting calibration is defined as:
\begin{equation}
    \ZAGN = (-1.00\pm0.09)P+(0.036\pm0.003)R_{23}+(8.80\pm0.06)
    \label{eq:dors21_calib}
\end{equation}
\noindent

\noindent    where $\ZAGN=12+\log(\rm O/H)$. 

The oxygen abundances used to derive the above calibration do still rely on photoionization models, though indirectly.  In particular, when only one auroral line is detected (in the case of \citealt{Dors2021}, the \OIIIauroral\ line), the temperature for part of the nebula is measured, and abundance estimates of the ionization species found within that region is well determined.  Yet, to derive the total oxygen abundance, one needs to account for all ionization species that exist in regions with different effective temperatures.  The temperature of these other regions must, therefore, be inferred using other methods. \cite{Dors2020b} created a grid of photoionization models to derive the relationship between the temperature of the high ionization region, $t_3$ (where oxygen is doubly ionized, traced by \OIIIall\ and \OIIIauroral), and the low ionization region, $t_2$ (where oxygen is singly ionized).  To derive a relation between $t_2$ and $t_3$ suitable for the NLR, \cite{Dors2020b} created a grid of models using ionizing spectra with power-law slopes of $\PLalpha = -0.8,-1.1,-1.4$, ionization parameters over the range $-3.5$ to $-0.5$ (again, defined at the inner radius of the ionized cloud), and metallicities over the range $0.2<\ZAGN/\Zsun<2.0$, and for a range of different electron densities from 100 to 3000 cm$^{-3}$. In this work, they follow  the relation $\log(\textrm{N/O}) = 1.29[\logOH]-11.84$ to assign a given nitrogen abundance. They also fit a relation between $t_2$ and $t_3$ for the whole grid, as well as for different electron densities.  This relation may therefore be reasonable for `typical' NLRs (if the grid presented in \cite{Dors2020b} is appropriate for `typical' NLRs), but individual galaxies will have very different $t_2/t_3$ values, as demonstrated by the wide spread in $t_2/t_3$ for their fiducial grid (their figure~5).  Treatment of depletion onto dust grains is not referred to in their publications.

\subsubsection{P\'erez-Montero et al. 2019}

\cite{Perez-Montero2019} extended the Bayesian-like \HIIChimistry\ code\footnote{This code establishes a Bayesian-like comparison between the predictions of the lines emitted in the ionized gas from a grid of photoionization models covering a large range of input parameters. In this comparison, the code does not assume any fixed relation between secondary and primary elements.} \citep{Perez-Montero2014} to include characterization of the NLR gas using a grid of photoionization models produced with \cloudy\ v.17.01.  The incident ionizing radiation consists of two components: that characterizing the `big blue bump' [the curved component peaking at $\log(\nu/\txn{Hz})\sim13.5$ in Fig.~\ref{fig:accretion disc}], and a power-law slope between 2keV (6\AA) and 5eV (2500\AA) of $\alpha_\txn{ox}=-0.8$ (they also compare to a model with $\alpha_\txn{ox}=-1.2$); power-law emission with $\alpha_x=-1$ representing the non-thermal X-ray emission.  We show the incident radiation in Fig.~\ref{fig:accretion disc}.  \cite{Perez-Montero2019} 
add dust within the NLR and model the gas as homogeneously-distributed with filling factor 0.1 and hydrogen density, $\nH=500\,\textrm{cm}^{-3}$.
All element abundances are assumed to scale to solar, except for nitrogen, which is an additional free parameter.  

\HIIChimistry\ uses standard line ratios to derive physical parameters, and works by first constraining N/O using the ratios between \NII\ and \OII\ and between \NII\ and \SII, before constraining metallicity and ionization parameter.  The chemical composition of the gas is characterized by the total oxygen abundance, which covers the range $6.9<\logOH<9.1$, $-2<\log(\textrm{N/O})<0$ and the released version of the code covers $-2.5<\logUsAGN<-0.5$.  The ionization parameter in these models is defined the same way as for \beagle\ (following description in section 3.1 of \citealt{Perez-Montero2014}).  The models consider default grain properties and relative abundances, and all elements but Nitrogen are scaled to the solar values given by \citet{Asplund2009} considering the \cloudy\ default depletion factors. Finally, we note that the code is not able to disentangle the contributions of star formation and NLR to the lines.

\subsubsection{Thomas et al. 2018}


\NebularBayes, presented in \cite{Thomas2018NebularBayes}, is the only code, other than \beagle, which simultaneously fits the star-forming and NLR contributions to nebular emission.  The NLR models are based on physically motivated prescriptions for the incident radiation produced by the accretion disc \citep{Thomas2016}.  The accretion disc model has three parameters: the energy of the peak of the accretion disc emission, E$_{peak}$; the photon index of the inverse Compton scattered power-law tail, $\Gamma$ (non-thermal X-ray radiation); and the proportion of the total flux that goes into the non-thermal tail, p$_{NT}$. The last two parameters are fixed to the fiducial values of $\Gamma=2.0$ and p$_{NT}=0.15$.

The NLR grid included in \NebularBayes\ is calculated for constant pressure models using \mappings\ v5.1 \citep{Sutherland2017}. 
This is different from the constant density models produced for the other works. \cite{Groves2004} demonstrated the large difference in coverage of the BPT plane that dusty isobaric models introduce. In particular, for the optical line ratios explored in this paper, their isobaric dusty model curves also better sample the parameter space covered by the data than their dust-free, constant density models. This is due to the absorption of ionizing photons by dust, which increases
the radiation pressure (and hence density), especially at high values of the ionization parameter.
The four parameters of the NLR that can be derived using \NebularBayes\ are: oxygen abundance of the gas; ionization parameter, \logUinAGN, defined at the inner edge of the ionized gas; peak of the ionizing radiation, \Epeak; and pressure.

\cite{Thomas2018} demonstrates \Epeak\ cannot be constrained when fitting to \OII\, 
\NeIII\,	\Hbeta\, \OIIIauroral\, \OIIIall\, \OI\, \HeI\, \Halpha\, \NII\, and \SII\ (those available in the SDSS DR7 MPA JHU measured line fluxes), but find that values in the range 40--50 eV give reasonable fractions\footnote{Higher $\Epeak=45$ values give a very low fractional contribution of NLR to \Hbeta\ line flux, whereas the values they choose span 0.1-1.0 along the AGN branch in the BPT diagrams.} of NLR contribution, and so fix this parameter to $\Epeak=45$ eV.

The grid of models includes 12 oxygen abundances in the range $\rm -1.7 \: \leq \: log(O/H) \:  \leq \: 0.54$, using the oxygen abundance scaling from \citet{Nicholls2017}. Depletion onto dust grains is considered based on iron being $97.8\%$ depleted. The ionization parameter at the inner edge of the nebula can take 11 uniformly spaced values in the range $-4.2\leq \logUinAGN \leq -0.2$. The initial gas pressure samples 12 values in the range $4.2 \leq \log P/k (\rm cm^{-3} K) \leq 8.6$ and values of \Epeak\ sample six values in the range $-2.0 \leq \log \Epeak (\rm keV) \leq -0.75$.

\subsection{Comparison to other works and data}\label{section:generalcomparison}

\begin{figure*}
\centering
  \subfigure[]{\includegraphics[width=\columnwidth]{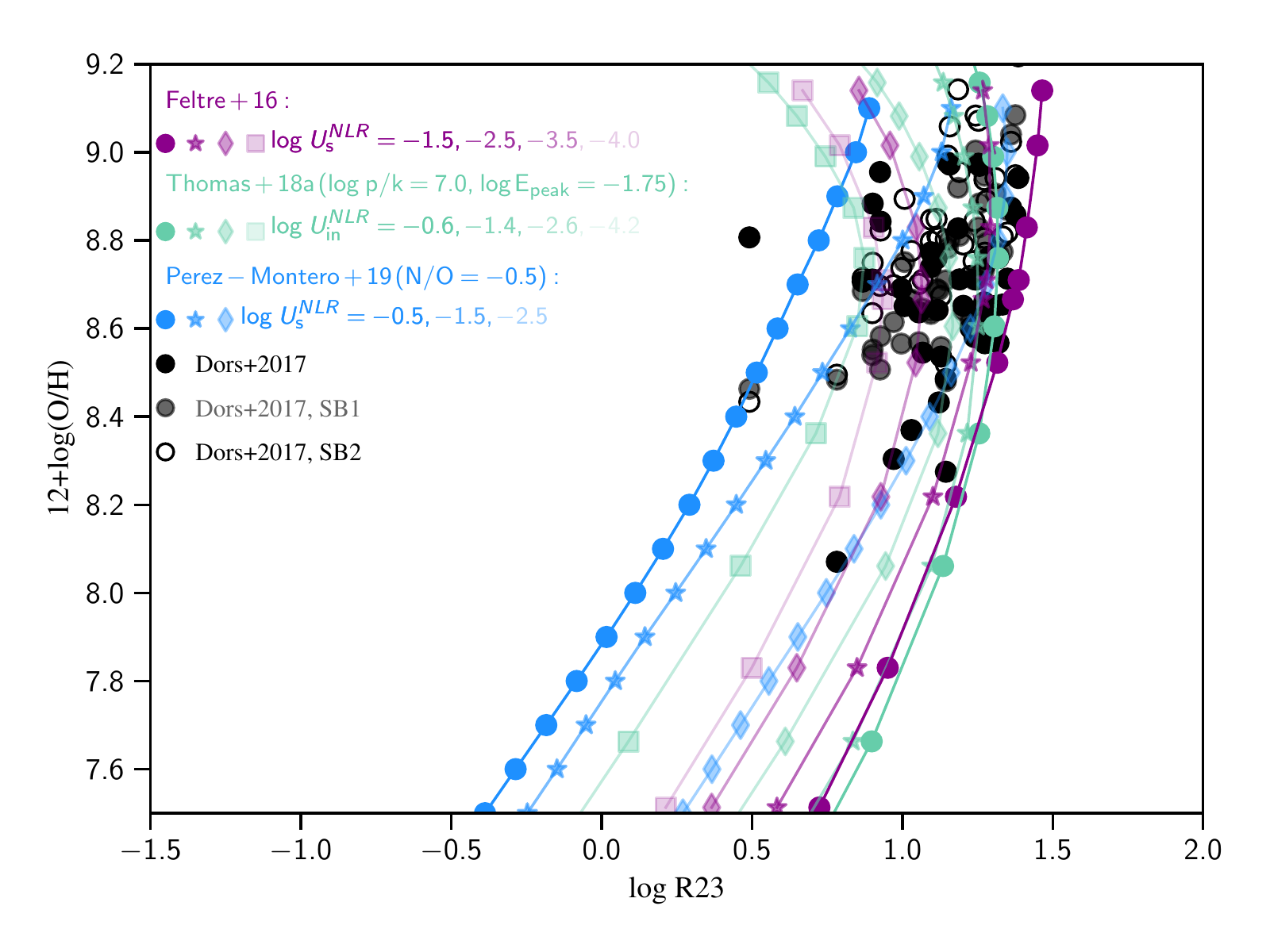}}
  \subfigure[]{\includegraphics[width=\columnwidth]{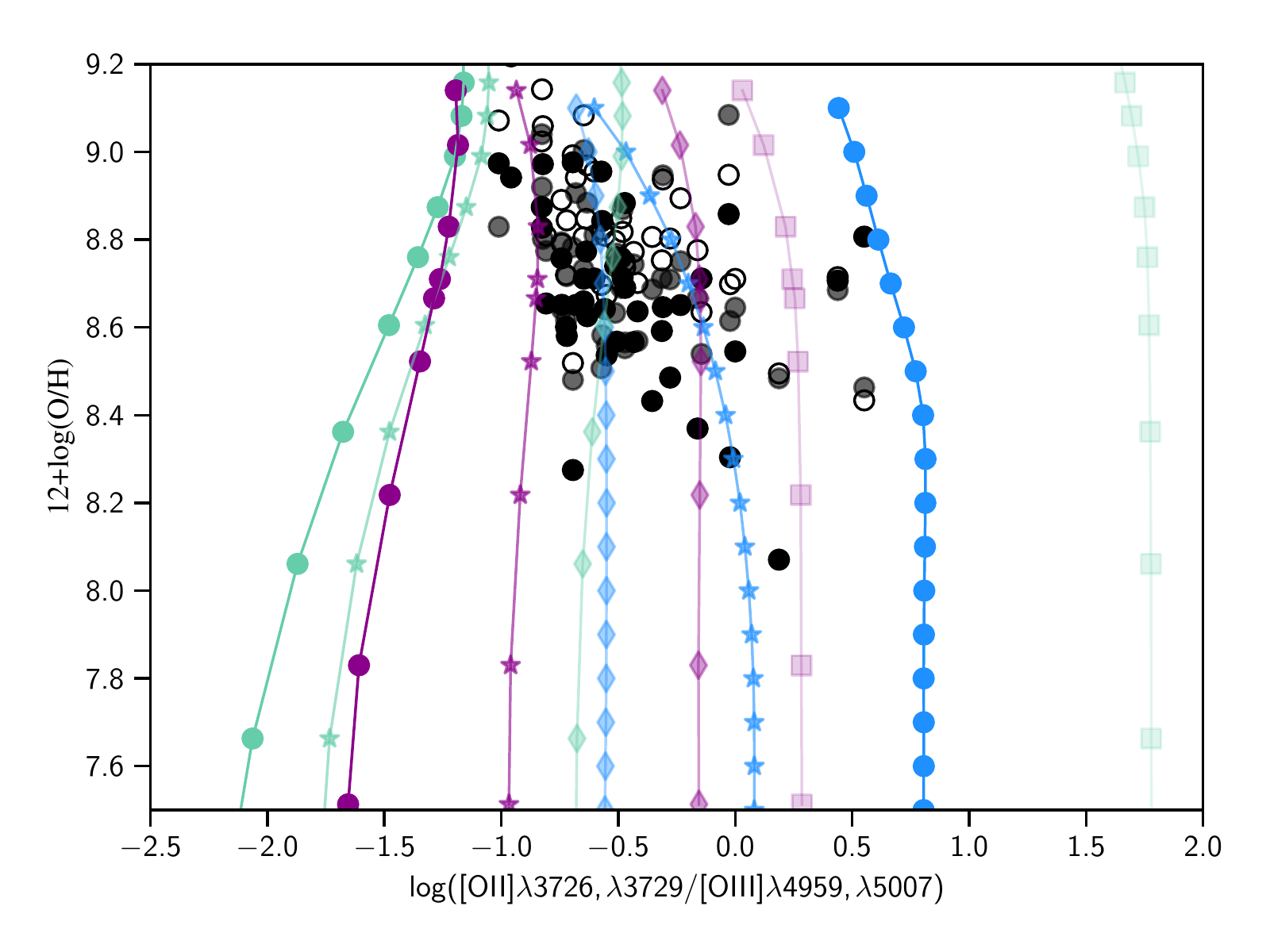}}
\caption{Gas-phase oxygen abundance (\logOH) versus different emission line ratios, log(R23) (defined in equation \protect\ref{eq:R23}) and log(\OII/\OIIIall).  In these figures we show the coverage of different NLR emission line models via lines and symbols with a range of metallicity and ionization parameters.  The symbols and lines range from fainter to darker colours from lower to higher \logUsAGN\ or \logUinAGN\ respectively, as indicated in the legend. The black, grey and open circles show the measured emission line ratios and inferred NLR \logOH\ values for a set of type-2 AGNs collated in \protect\cite{Dors2017a}.  Black circles display the \logOH\ estimates from \protect\cite{Dors2017a}, while the grey and open circles display the \logOH\ estimates calculated with the two \protect\cite{Storchi-Bergmann1998} NLR calibrations, defined in equations \ref{eq:sb1} and \ref{eq:sb2}.}
\label{fig:logOH_vs_ELratios}
\end{figure*}

We start by comparing our NLR-emission models (F16) to those of \cite{Thomas2018NebularBayes} and \cite{Perez-Montero2019}, as well as observed line ratios of 44 observed Seyfert II galaxies collated by \cite{Dors2017a}. Fig.~\ref{fig:logOH_vs_ELratios} shows gas-phase \logOH\ against two different emission line ratios: R$_{23}$ (defined in equation~\ref{eq:R23}) and $\log(\OII/\OIIIall)$. 
The lines and matching coloured symbols show the coverage of different models in these plots, while the black and grey symbols show measured line ratios and estimated abundances for the observed galaxies. For each of the observed galaxies, we show three different abundance estimates:  those from \cite{Dors2017a} (filled black circles) which were estimated by comparing directly to \cloudy\ photoionization models; 
and those derived with each of the \cite{Storchi-Bergmann1998} calibrations, with filled grey (open black symbols) showing the measurements obtained with the calibration in equation~\ref{eq:sb1} (\ref{eq:sb2}).  As noted by \cite{Storchi-Bergmann1998}, the second calibration (equation~\ref{eq:sb2}) leads to systematically higher oxygen abundance estimates than the other two methods.  

The models in Fig.~\ref{fig:logOH_vs_ELratios} span a range of ionization parameters ranging from high to low values with dark-to-faint symbols.  For the F16 models, we set $\PLalpha=-1.7$.  \cite{Thomas2018} fix E$_{peak}$ to 45 eV, which is somewhat higher than the peak incident radiation within our models. We investigated values in the range $8<\Epeak/\txn{eV}<56$ but in order not to crowd the plot too much, we show the results for a central value within this range of $\sim17.8$ eV. We also explored the full pressure range $4.2<\log(\txn{P/k})<8.6$, but plot those with the central value of $\log(\textrm{P/k})=7$. Then, since the \cite{Perez-Montero2019} model grid encompasses a wide range of N/O abundances, we fix to a value of log(N/O)=$-0.5$ (the same considered as the `standard' one by \cite{Perez-Montero2019}), although the diagrams in Fig.~\ref{fig:logOH_vs_ELratios} are not very sensitive to the value chosen. 

Fig.~\ref{fig:logOH_vs_ELratios} (a) shows the oxygen abundance versus R23, which is defined in equation~\ref{eq:R23}.
The F16 models cover the full range of the observations, with the exception of a few objects with the lowest values of R23. 
The \cite{Perez-Montero2019} models provide good coverage of the observations with low values of R23, but provide only marginal coverage of the highest R23 values.  We display the \cite{Thomas2018} models with a single value of \Epeak, although have tested a range from 8 to 56 eV, finding that the models cover the full range of observed values,  except for those few observations with the lowest R23.   It is important to note that the abundance estimates of the observed objects plotted are not objectively more secure than other abundance estimates presented in this work.  They are subject to their own modelling assumptions, so we focus mostly on the coverage of the plotted models over the emission line ratio space.  

Fig.~\ref{fig:logOH_vs_ELratios} (b) shows the gas-phase oxygen abundance for the same models and data as in Fig.~\ref{fig:logOH_vs_ELratios} (a), but plotted against $\OII/\OIIIall$.  The \cite{Thomas2018NebularBayes} and \cite{Perez-Montero2019} models do not include [O{\sc iii}]$\lambda4959$, which we assume to have 1/3 the flux of \OIII.  We see that for the range of ionization parameters plotted, our models  cover the parameter space spanned by the observations for all but the highest $\OII/\OIIIall$ ratios.  
The \cite{Perez-Montero2019} models cover the highest $\OII/\OIIIall$ values, but observations with $\OII/\OIIIall\lesssim-0.5$ are not covered.  The \cite{Thomas2018NebularBayes} models cover the full range of $\OII/\OIIIall$ ($\lesssim$-0.5) from the observations.

\begin{figure}
  \centering
  \includegraphics[width=\columnwidth]{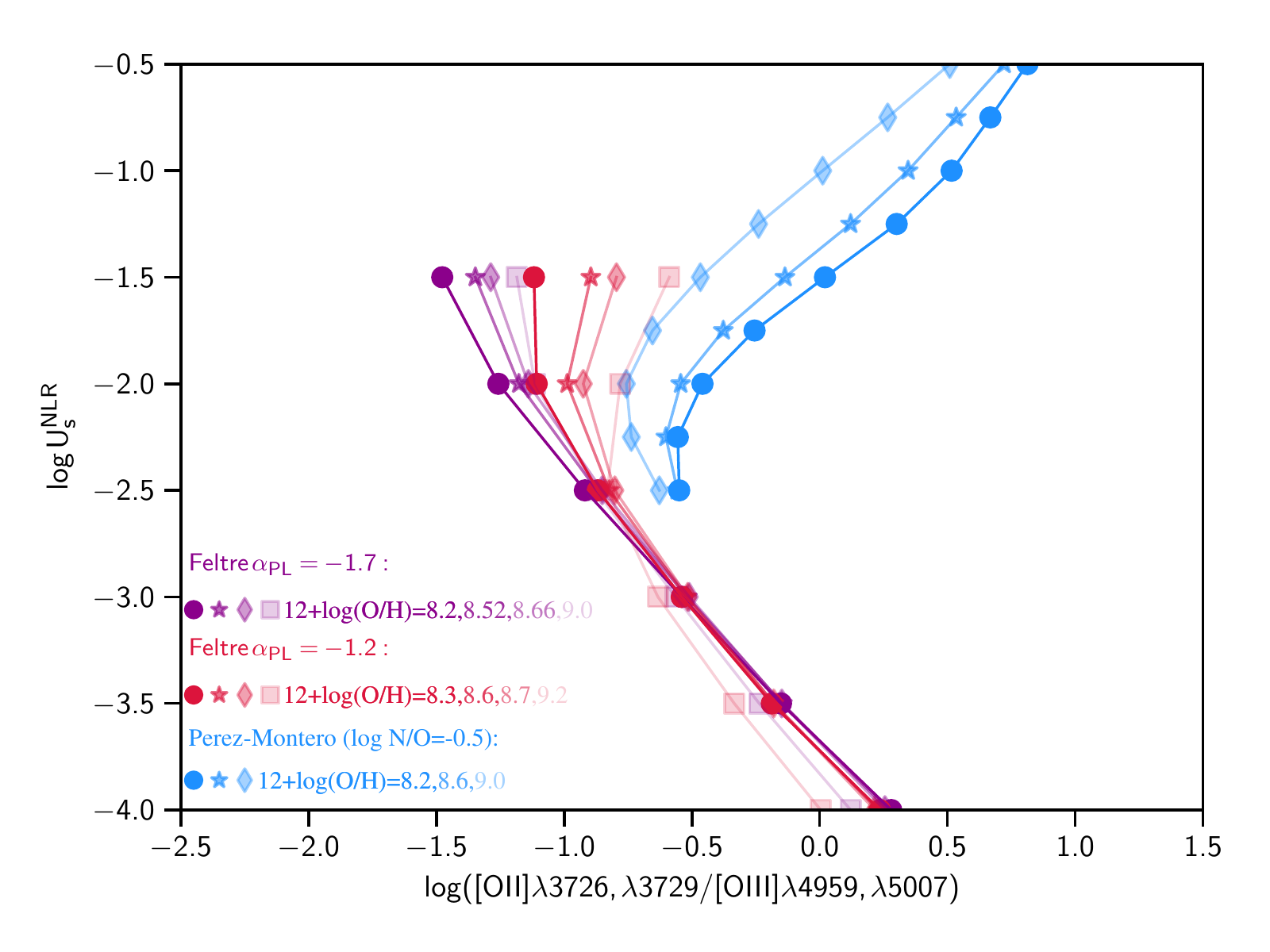}
  \caption{\logUsAGN\ or \logUinAGN\ versus \OII/\OIIIall\ for the F16 and \protect\cite{Perez-Montero2019} models for a range of different \logOH\ values.  We show the F16 models with two values of $\PLalpha$: -1.2 (red) and -1.7 (violet), as indicated in the legend. 
   }
  \label{fig:logU_O2O3}
\end{figure}

Both panels of Fig.~\ref{fig:logOH_vs_ELratios} show that the \cite{Perez-Montero2019} models give the opposite dependence of R23 and \OII/\OIIIall\ on \logUsAGN\ to those of F16 and \cite{Thomas2018NebularBayes}.  To understand this, we plot \logUsAGN\ as a function of \OII/\OIIIall\ in Fig.~\ref{fig:logU_O2O3}.  \cite{Perez-Montero2019} chose to publish only the highest \logUsAGN\ models, as the behaviour between \logUsAGN\ and \OII/\OIII\ is double-valued in their models.  We do not see the same behaviour in the F16 models with $\PLalpha=-1.7$, but when we plot our models with shallower $\PLalpha=-1.2$, we start to see a similar behaviour at the highest \logUsAGN\ and \logOH\ values (red symbols).  The strong increase in \OII/\OIII\ with \logUsAGN\ of the \cite{Perez-Montero2019} models is therefore presumably due to the choice of a hard incident ionizing spectrum (Fig.~\ref{fig:accretion disc}). 
This is also likely the reason for the moderate offset to higher \OII/\OIII\ values at given \logUsAGN\ compared to the F16 models (given that the partially ionized zone will be more extensive).  

\begin{figure*}
  \centering
  \includegraphics[width=7in]{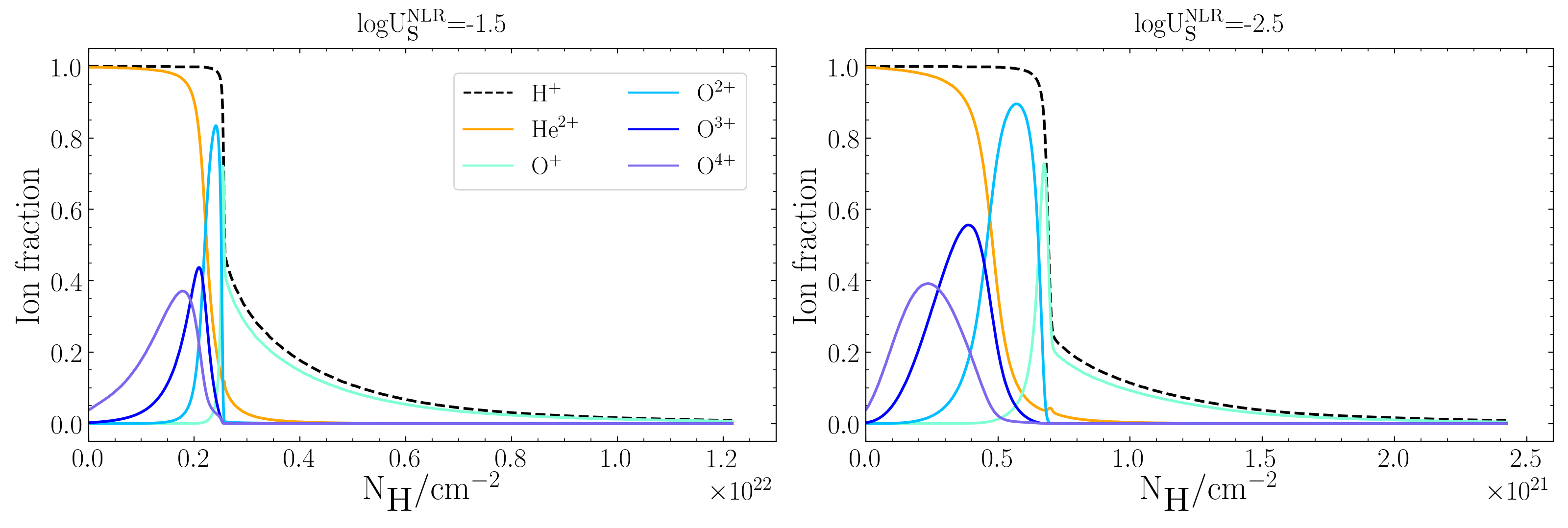}
  \caption{The ion fractions of $\rm{H}^{+}$ (black), $\rm{He}^{2+}$ (orange), $\rm{O}^{+}$ (turquoise), $\rm{O}^{2+}$ (sky blue), $\rm{O}^{3+}$ (dark blue) and $\rm{O}^{4+}$ (purple) as a function of the total hydrogen column density. The NLR models have $\ZAGN=0.030$, $\PLalpha=-1.2$ and $\logUsAGN=-1.5$ (left) and -2.5 (right).
  }
  \label{fig:ionization_structure}
\end{figure*}

Including only the higher part of the \logUsAGN\ fork, the \cite{Perez-Montero2019} models will measure an opposite \logOH\ versus \logUsAGN\ to that measured with the \cite{Thomas2018} or F16 models, which will in turn provide systematically different oxygen abundances.  The data used in the \cite{Perez-Montero2019} fitting could not distinguish which branch of \logUsAGN\ versus \OII/\OIIIall\ the data inhabit.  It would clearly be beneficial to better constrain the ionization state of NLRs in observed type-2 AGN to motivate the range of \logUsAGN\ the models must span.   Fig.~\ref{fig:ionization_structure} displays the ion fraction for a number of different species (hydrogen, helium and oxygen) from the F16 models with $\PLalpha=-1.2$. The ion fraction are shown as a function of the total hydrogen column density, $\rm{N}_{\rm{H}}(R)=\nHAGN \epsilon R$. The high energy photons from hard ionising radiation of the accretion disc penetrates further through the gas than lower energy photons, creating a partially ionized region. We might expect \OII/\OIII\ to decrease with increasing \logUsAGN.  However, as explained in \cite{Perez-Montero2019}, and shown clearly in Fig.~\ref{fig:ionization_structure}, for high ionization parameters and hard ionizing radiation, this is not the case as at small radii within the Str\"{o}mgrem radius, \OIII\ is lost to higher ionization states of oxygen such as $\rm{O}^{3+}$ and $\rm{O}^{4+}$, while $\rm{O}^{+}$ is still present in the partially ionized zone.   There are no visible emission lines from these higher ionization states in the wavelength range of SDSS.  However, He\,{\sc ii} has a similar ionization energy 
to \textsc{Oiv} (54.4eV versus 54.9eV), and the SDSS spectra provide coverage of the \HeII\ recombination line.

\subsection{Nebular \HeII\ fluxes to constrain the ionization state of the NLR in type-2 AGNs within SDSS}\label{section:HeII}

We measure the nebular \HeII\ fluxes for 463 confirmed type-2 AGNs presented in \cite{Dors2020} from DR7 SDSS spectra \citep{Abazajian2009}.  \HeII\ is not delivered in the MPA-JHU group\footnote{https://wwwmpa.mpa-garching.mpg.de/SDSS/DR7/} distribution of measured line fluxes used in \cite{Dors2020}, since measurements of this line can be complicated by stellar wind features and broadening. We measured \HeII\ using the python package \textsc{emcee} \citep{emcee}.  Since we wish to gain constraints on the nebular (narrow-line) contribution to the \HeII\ line, we tie the width of the line to the width of \OIII\ in the same spectra, by simultaneously fitting to the two lines.  This effectively uses information from the higher S/N \OIII\ line to provide a strong prior on the shape of \HeII\ if it is to arise in the same physical region.  We define continuum regions on either side of the two lines and subtract a linear fit to the continuum before performing the line fits.  In the case where no obvious nebular \HeII\ emission is present, the tied fitting will provide firm constraints on the limits on the \HeII\ flux with the delivered uncertainties.  We report the measured line fluxes and line widths in Table~\ref{tab:HeII_measurements}. 

\begin{table}
\centering
  \caption{\HeII\ line fluxes measured for the sample of type-2 AGNs presented by \protect\cite{Dors2020}. Full sample will be available online.} 
  \begin{tabular}{|c|c|c|}
  \hline
  MJD-PLATE-FIBER & \HeII\ flux & width\\
                  & (/10$^-16$ erg s$^{-1}$ cm$^{-2}$) & (/\AA) \\
  \hline
51882-0442-061 & $0.826\pm0.270$ & $2.36\pm0.04$\\
52056-0603-545 & $0.209\pm0.145$ & $2.03\pm0.23$\\
51691-0340-153 & $0.258\pm0.192$ & $4.19\pm0.86$\\
52079-0631-387 & $0.231\pm0.152$ & $1.95\pm0.05$\\
52017-0516-232 & $1.01\pm0.26$   & $3.03\pm0.12$\\
51915-0453-002 & $1.17\pm0.30$   & $3.61\pm0.08$\\
52049-0618-535 & $0.261\pm0.184$ & $3.31\pm0.23$\\
52000-0335-428 & $0.135\pm0.116$ & $1.83\pm0.33$\\
  \hline
\end{tabular}
\label{tab:HeII_measurements}
\end{table}

\begin{figure*}
  \centering
  \includegraphics[width=3.4in]{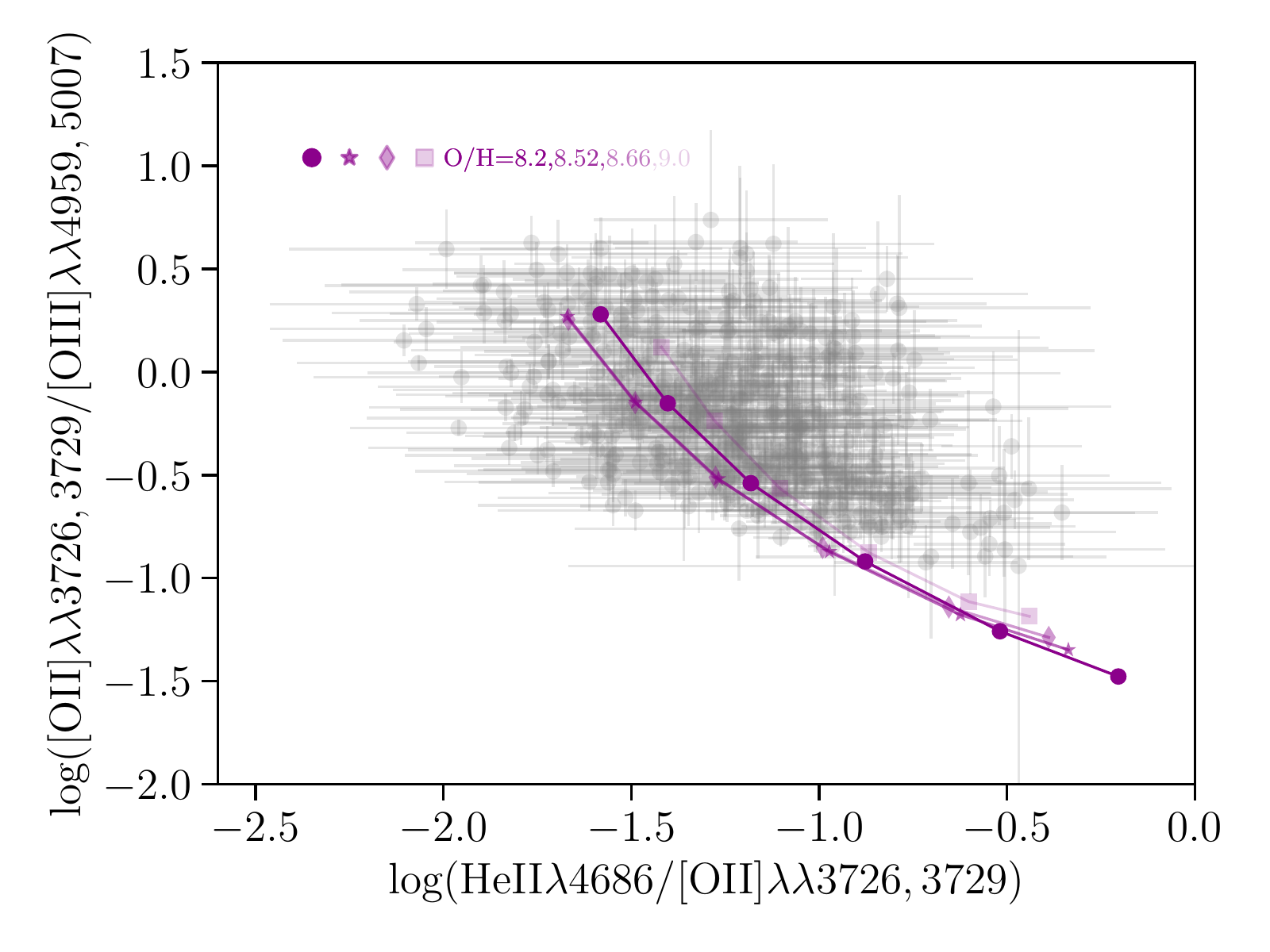}
  \includegraphics[width=3.4in]{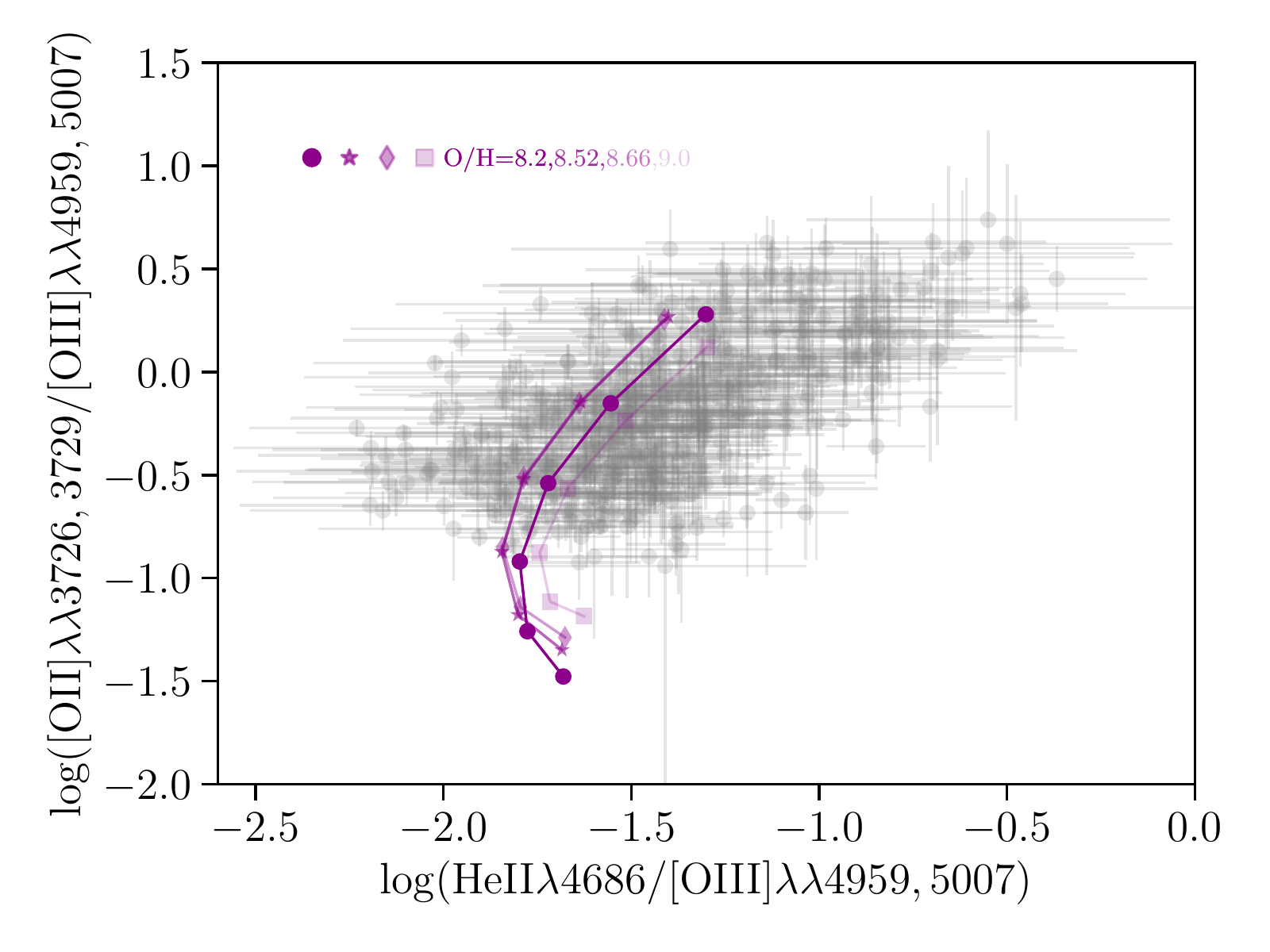}
  \caption{Assessing the ionization state of the gas within the type-2 AGN sample from \protect\cite{Dors2020}. With our measurements of the \HeII\ line from the SDSS spectra of these objects (see text for details), we compare log(\OII/\OIIIall) versus log(\HeII/\OII) [log(\HeII/\OIIIall)] to the F16 models with \PLalpha=-1.7 in the left [right] panels.  The highest \logUsAGN\ models have low \OII/\OIIIall\ ratios. The data shows a clear trend of decreasing \OII/\OIIIall\ with increasing \HeII/\OII\ or decreasing \HeII/\OIIIall, as shown by the F16 models.}
  \label{fig:HeII_measurements}
\end{figure*}

We show \OII/\OIIIall\ against \HeII/\OII\ (left) and \HeII/\OIIIall\ (right) in Fig.~\ref{fig:HeII_measurements}, with the F16 models for comparison.  For this figure, we used the MPA-JHU line fluxes for \OII\ and \OIIIall.  The sample of type-2 AGNs clearly follow the general trend of the F16 models.  Unfortunately the models of \cite{Perez-Montero2019} do not include \HeII, so we cannot compare to their predicted ratios directly.  If \OII/\OIIIall\ were increasing with \logUsAGN\ (as for the \citealt{Perez-Montero2019} models in Fig.~\ref{fig:logU_O2O3}), we would expect \HeII/\OII\ to increase with \OII/\OIIIall, as well as much higher values of \HeII/\OII\ for the observed galaxies. The observations show low values of \HeII/\OII\ and a negative trend between \OII/\OIIIall\ and \HeII/\OII\ (Fig.~\ref{fig:HeII_measurements}, a), suggesting that the set of observed type-2 AGNs have significantly lower ionization parameter than covered by the \cite{Perez-Montero2019} models. 

\subsection{Comparison of the nitrogen and sulfur emission lines}\label{section:nitrogen}

\begin{figure}
  \centering
  \includegraphics[width=\columnwidth]{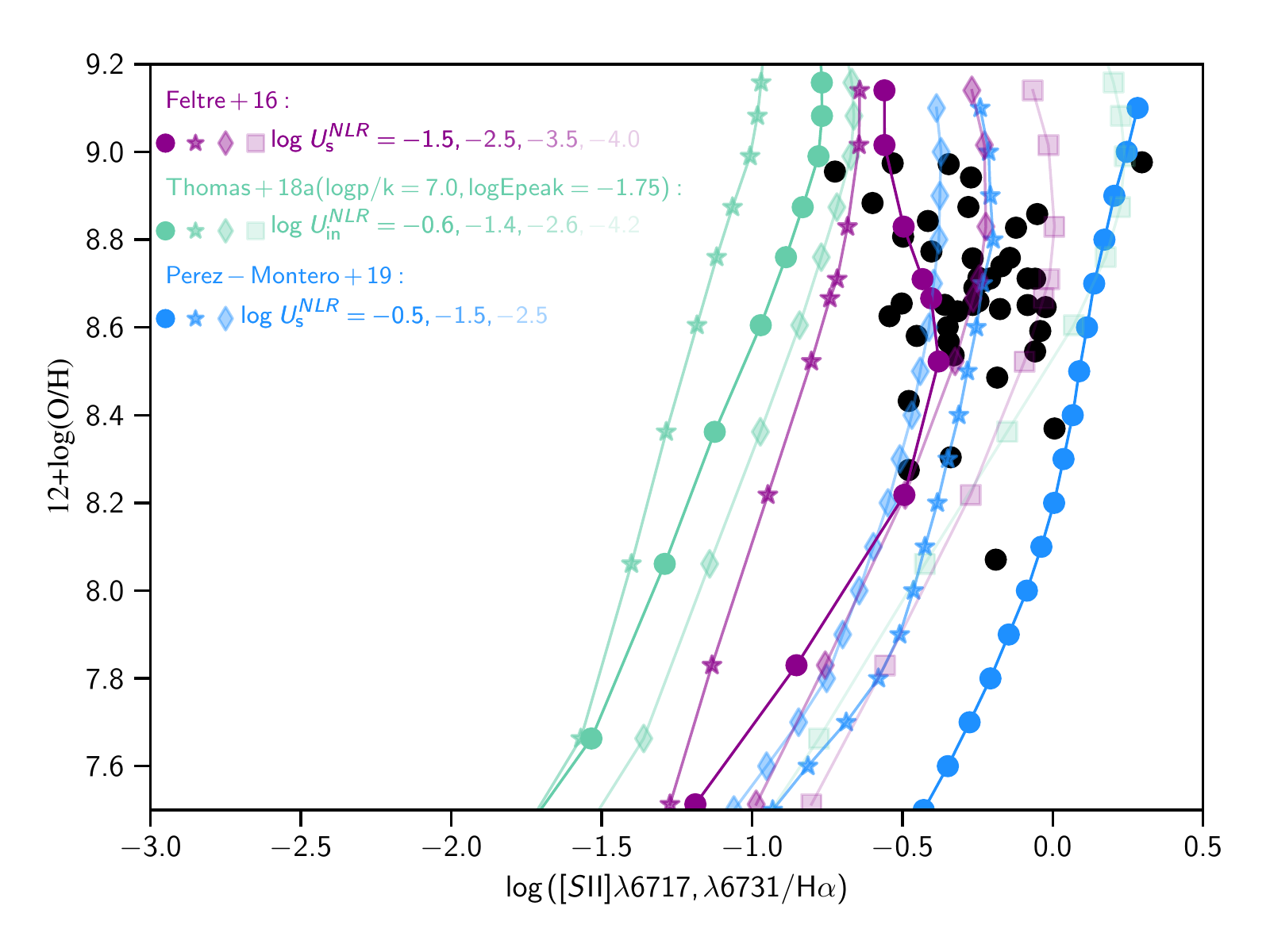}
  \caption{Gas-phase oxygen abundance (\logOH) as a function of \SII/\Halpha\ for the same models as in Fig.~\ref{fig:logOH_vs_ELratios}.}
  \label{fig:SIIHalpha}
\end{figure}

Because the \SII\ doublet is routinely used to derive the gas density in photoionized regions, we have also compared the predictions for this doublet from the different NLR models and data presented in the previous sections. In particular, we show in Fig.~\ref{fig:SIIHalpha} the gas-phase oxygen abundance as a function of the ratio \SII/\Halpha. In this case, we only show the data with the predictions for \logOH\ from \citealt{Dors2017a} in order to not crowd the figure too much. The \citealt{Perez-Montero2019} models reproduce the data except for the points with the highest oxygen abundances and lowest \SII/\Halpha\ ratios. The coverage of the F16 and \citealt{Thomas2018NebularBayes} is very similar and almost complete except for three objects with high \SII/\Halpha\ ratios. We have noted (not shown here) than for lower values of \PLalpha\ F16 models cover better this region of the parameter space.

Abundance studies of galactic and extragalactic \HII\ regions find that nitrogen has a primary origin for the low-metallicity regime (\logOH $\lesssim 8.2$) and a secondary one for the high-metallicity regime \citep[e.g.][]{Mouhcine2002,Perez-Montero2009}. \citealt{Groves2004} find that the nitrogen abundance from the primary and secondary origin can be related to that of oxygen using an equation of the form Equation~\ref{eq:Nitrogen}. To capture the richness of the variation of the N abundance as a function of O, we have expanded the grid of NLR models.  The fiducial grid has a N/O dependence on the oxygen abundance following: 
\begin{equation}\label{eq:NOrel}
    \rm{log}(\rm{N/O})_{\rm{tot}} \approx \rm{log} \{ 10^{-1.732}+10^{[\rm{log(O/H)_{\rm{tot}}}+2.19]}\} +C
\end{equation}
with $\rm C=-0.25$.  Now we are parametrizing N/O in terms of the \textit{total} abundances, $\rm{(N/O)}_{\rm{tot}}$ (i.e. gas + dust) and its dependence on the \textit{total} oxygen abundance, $\rm{(O/H)}_{\rm{tot}}$ (gas + dust).  In expanding the grid we add two new values of the normalisation, $C$, with $\rm C=0$ (the \citealt{Nicholls2017} relation) and $\rm C=0.5$ (the upper limit).
See Plat et al. (in prep.) for the details of how these models have been computed. 

We test the nitrogen abundance in the different grid of models using the N2 and O3N2 line ratios, defined as follows:

\begin{equation}
    {\rm N2}=\log \left(\dfrac{\NII}{\Halpha}\right)
    \label{eq:N2}
\end{equation}
\noindent


\begin{equation}
    {\rm O3N2} = \log \left(\frac{\OIII}{\Hbeta}\frac{\Halpha}{\NII}\right)
    \label{eq:O3N2}
\end{equation}
\noindent

\begin{figure*}
\centering
  \subfigure[]{\includegraphics[width=\columnwidth]{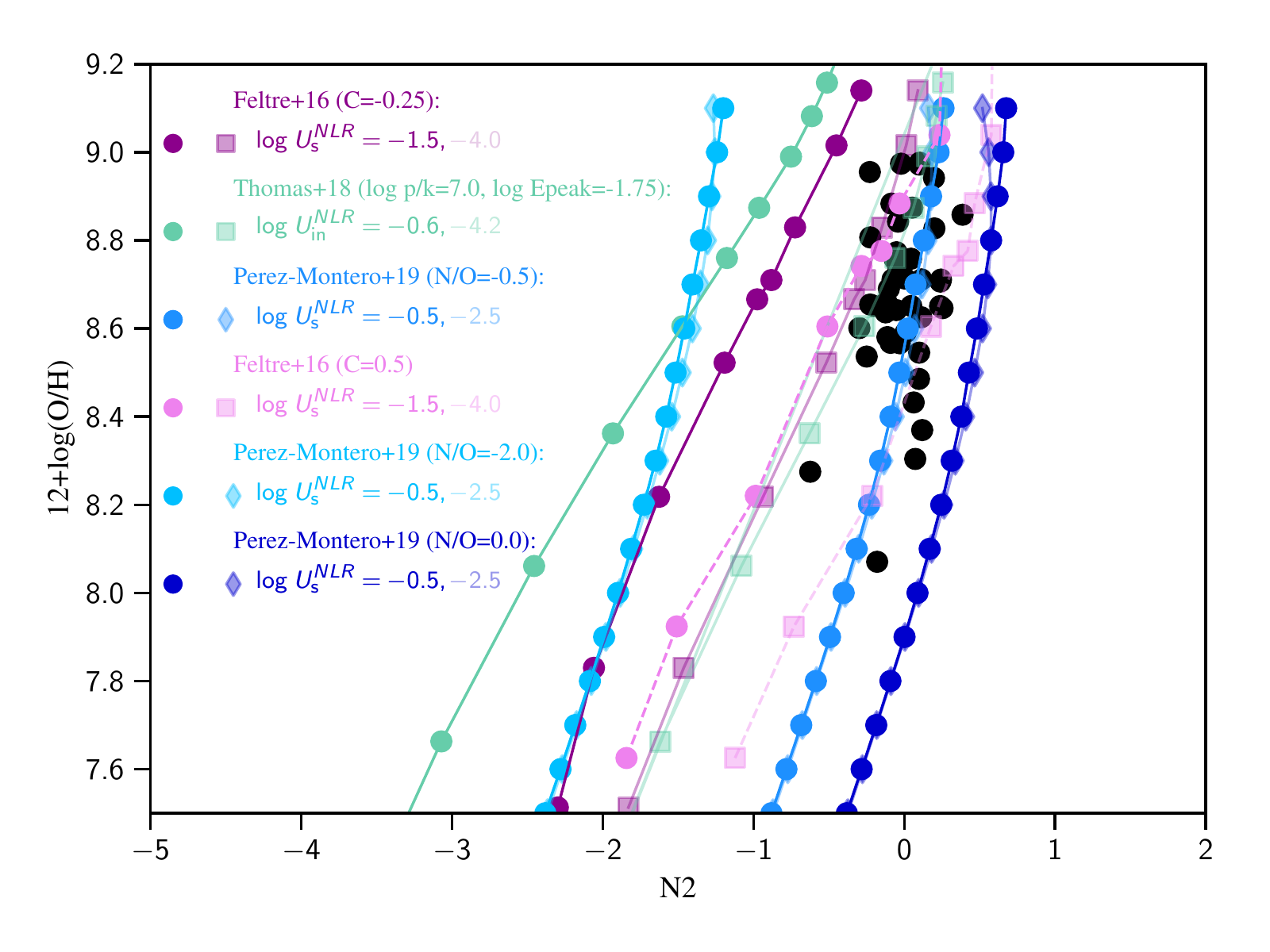}}
  \subfigure[]{\includegraphics[width=\columnwidth]{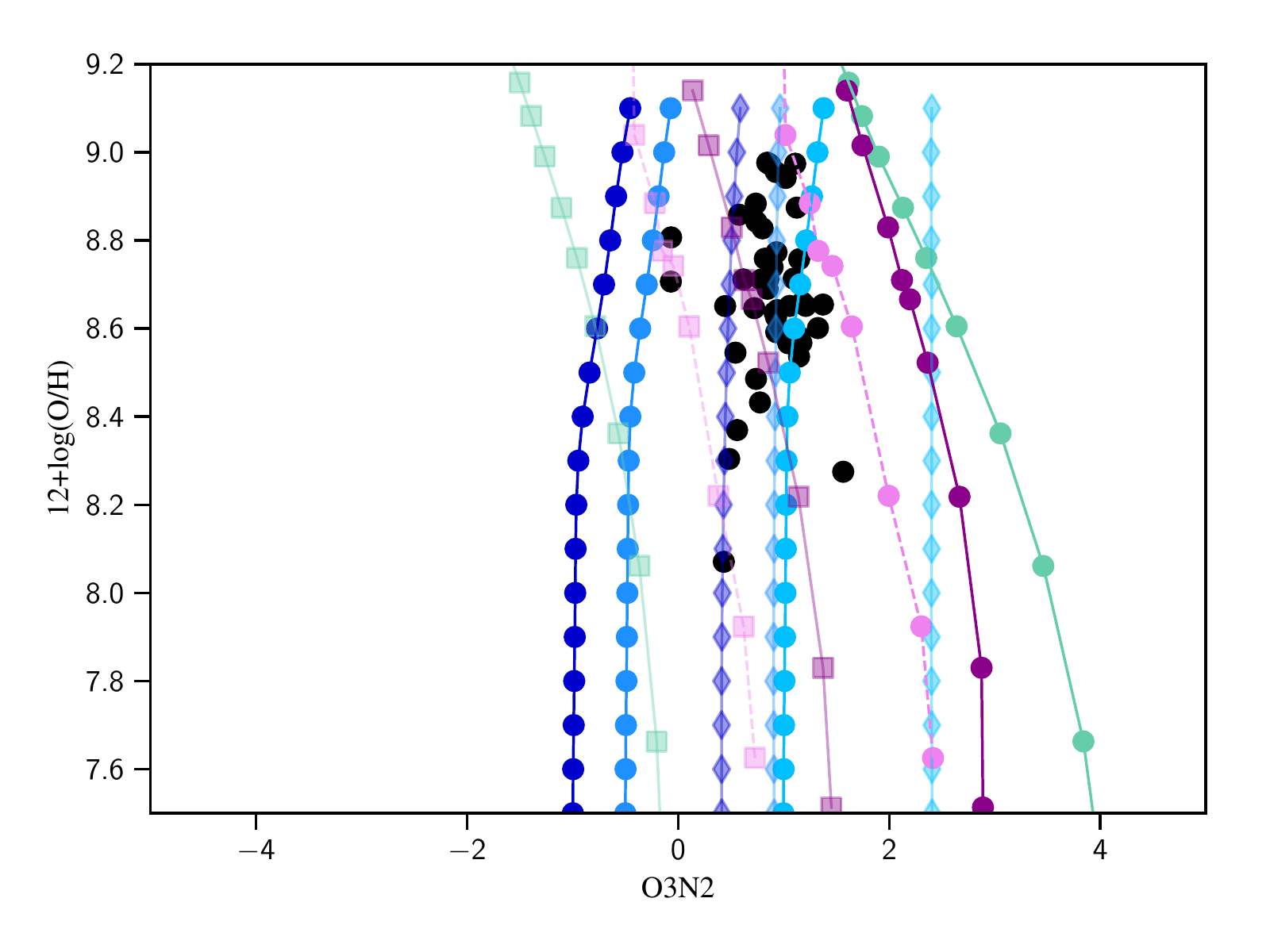}}
\caption{Gas-phase oxygen abundance (\logOH) as a function of N2 (a, see equation~\ref{eq:N2}) and O3N2 (b, see equation~\ref{eq:O3N2}) for the same models as in Fig.~\ref{fig:logOH_vs_ELratios} and different variations of the N/O relation: the \citealt{Gutkin2016} relation ($\rm C=-0.25$) and the upper limit from eq.~\ref{eq:NOrel} ($\rm C=0.5$) for the F16 models and the highest and lowest values for nitrogen abundance of \citealt{Perez-Montero2019} grid. The data points are from \citealt{Dors2017a}. } 
\label{fig:N2}
\end{figure*}

We show in Fig.~\ref{fig:N2} the gas phase oxygen abundance as a function of these line ratios. For completeness, we show F16 and \citealt{Perez-Montero2019} models with different N/O values. In particular, for the \citealt{Perez-Montero2019} models, we add the ones with the lowest and highest value of N/O (cyan and dark blue points) and for the F16 models, the ones with the highest ($\rm C=0.5$, pink points with dashed lines) and lowest ($\rm C=-0.25$ that of equation~\ref{eq:Nitrogen}) scaling factor of the N/O relations explored. We recall that nitrogen abundance is not a free parameter in the \citealt{Thomas2018NebularBayes} grid. Most likely due to the fact that nitrogen abundances is fixed, we see in Fig.~\ref{fig:N2}a how \citealt{Thomas2018NebularBayes} models are not able to sample the data points with the highest nitrogen abundances. However, increasing the N/O value allows for a full and almost full coverage of the data for \citealt{Perez-Montero2019} and F16 models respectively. In Fig.~\ref{fig:N2}b the three grids of models sample well the space of the diagram where the data lie. We note however, that in the case of \citealt{Thomas2018NebularBayes} models, for data points with low O3N2 values, the \logUsAGN\ parameter might be underestimated. This might be due to the lack of higher Nitrogen abundances that will tend to be compensated in the O3N2 ratio with weaker \OIII\ lines, and therefore, lower \logUsAGN\ values.

\beagle\ does not yet allow for sampling over N/O, though this comparison to observations shows that it is vital to adequately explain measured line ratios including Nitrogen.  It is beyond the scope of this paper to implement and test retrieval of parameters including the new N/O grid with our pedagogical approach, but this functionality will be implemented for future work.

\subsection{Comparing to the Dors 2021 $T_{\mathrm e}$-based empirical emission-line calibration}\label{section:comparingDors}

\begin{figure*}
  \centering
  \subfigure[]{\includegraphics[width=\columnwidth]{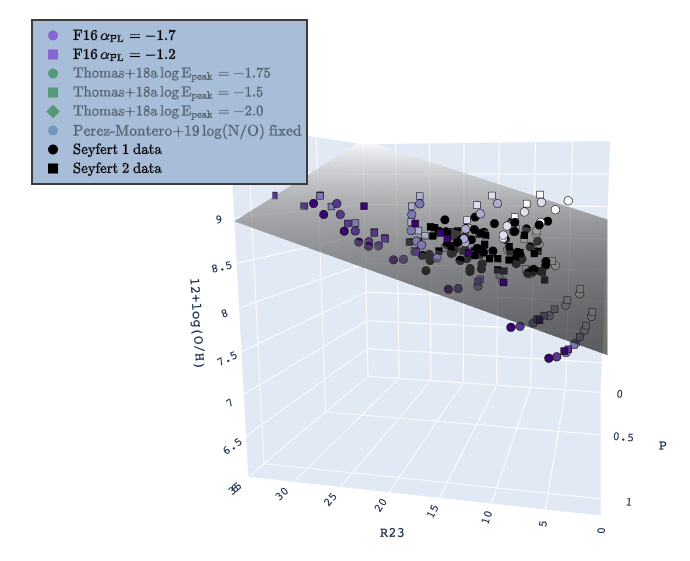}}
  \subfigure[]{\includegraphics[width=\columnwidth]{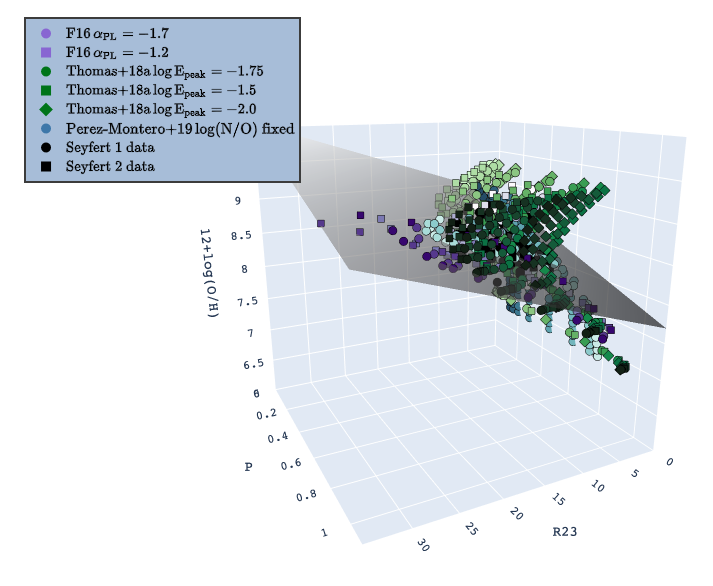}}
  \caption{\protect\cite{Dors2021} empirical $T_\textrm{e}$-based \logOH\ calibration, shown as the grey surface in \logOH--P--R23 space (where P and R23 are defined in Eqs.~\ref{eq:dors21_p} and \ref{eq:R23}, respectively).  Both panels show a sample of observed AGNs with \logOH\ estimated from the $T_\textrm{e}$ method by \protect\cite{Dors2021} (black points).  The F16 models are shown in panel (a) for two values of \PLalpha, and for a range of ionization parameters spanning $-4<\logUsAGN<-1.5$ with dark-to-light colours spanning high-to-low \logUsAGN\ values.  Panel (b) includes the \protect\cite{Thomas2018NebularBayes} and \protect\cite{Perez-Montero2019} models (see text for details).  The 3D version of these figures can be found at
  \url{https://chart-studio.plotly.com/~AlbaVidalGarcia/15/\#/}.}
  \label{fig:fig3Ddors21}
\end{figure*}

We plot the different theoretical models against the \cite{Dors2021} empirical emission-line calibration in Fig.~\ref{fig:fig3Ddors21}.  The calibration describes a surface in \logOH--P--R23 space (as described in Eqs.~\ref{eq:R23} and~\ref{eq:dors21_p} in Section~\ref{sec:Dors21}).  This provides a quick comparison of the models to $T_{\mathrm e}$-based NLR oxygen-abundance estimates without running full fits with \beagle.  We will present full \beagle\ fits to SDSS type-2 AGNs in a future paper in this series.  The black symbols display the $T_{\mathrm e}$-based gas-phase oxygen abundance measurements for the set of AGNs used to produce this empirical calibration, while the grey plane shows the empirical calibration.  

Fig.~\ref{fig:fig3Ddors21} (a) shows the F16 models plotted for a range of \logUsAGN, \logOH, and two values of \PLalpha\ ($-1.2$ and $-1.7$). For a given \PLalpha\ and \logOH, different colours represent different \logUsAGN\ values, with dark-to-light colours spanning high-to-low values.
We see that at very low metallicities our models sit below the plane describing the \cite{Dors2021} empirical relation.  Within the region occupied by the observed galaxies, the models agree well with the empirical relation until high metallicities, where the models curve up from the empirical plane.  The observations also preferentially scatter above the relation.  The behaviour of the models suggests this scatter to higher oxygen abundances is real, not simply intrinsic scatter about the empirical plane. Therefore, objects above the relation have higher oxygen abundances than can be explained by their position in the P-R23 plane, while the models can capture this behaviour. This behaviour is also observed in the \cite{Thomas2018NebularBayes} models (constant pressure models). The curvature from the plane is reminiscent to the effect of the addition of dust in NLR in \cite{Groves2004} as the change from constant density to constant-pressure models. However, our constant density models show a similar behaviour, indicating that the observed curvature might be due to the presence of dust, self consistently modelled in \cloudy.

The $\PLalpha=-1.7$ models do not cover all the observed data points in Fig.~\ref{fig:fig3Ddors21} (this might be best appreciated by viewing the interactive plot online, where different models can be removed from the plot for clarity \url{https://chart-studio.plotly.com/~AlbaVidalGarcia/15/#/}).  A range of $\PLalpha$ appears to be required to fully cover 
the data. $\PLalpha=-1.2$ covers the data well.  With our idealized grid exploration in Section~\ref{section:retrieval}, we found that the lines chosen for that analysis could not fully constrain $\PLalpha$, and that we recommend the user to set it to $\PLalpha=-1.7$ unless there is a significant improvement in the goodness of fit when \PLalpha\ is allowed to vary.  We defer the search of other lines that can constrain better this parameter to a later paper.

Panel (b) of Fig.~\ref{fig:fig3Ddors21} shows the same models, observations and empirical relation as shown in panel (a), but we include also the \cite{Thomas2018NebularBayes} and \cite{Perez-Montero2019} models.  In particular, for the \cite{Thomas2018NebularBayes}, we plot a range of $-4.2<\logUinAGN<-0.2$, with dark-to-light colours representing high-to-low values, and for three values of \Epeak, as indicated in the legend. In particular, we show the one taken as standard throughout this paper, -1.75 (green points) and a lower and a higher one, -2.0 (green diamonds) and -1.5 (green squares) respectively.  For the \cite{Perez-Montero2019} models, we plot a range of $-2.5<\logUsAGN<-0.5$ but we do not vary the N/O values because this variation has a negligible impact on the P and R23 values.  This figure shows that \textit{all} models curve off the empirical relation to high metallicities. 

This comparison to the \cite{Dors2021} empirical metallicity calibration shows the limitations of both the empirical diagnostic itself and our models.  The empirical diagnostic cannot capture the variation with metallicity above the plane. Our models require a range of \PLalpha\ to account for the data, a parameter we cannot constrain from the set of line ratios we tested. Our tests in Section~\ref{section:retrieval} further demonstrate how biased other NLR gas parameters become when allowing \PLalpha\ to vary.  However, below the empirical relation (lower metallicities), a region of the parameter space we expect to find at higher redshifts with missions such as \textit{JWST}, the difference between the two values of \PLalpha\ is less significant for the F16 models.  It is beneficial to use Te-based O abundance estimates, empirical diagnostics, and these types of sophisticated radiative transfer models together to tease out the limitations of each approach. Until $T_\textrm{e}$ measurements are possible for large samples of type-2 AGNs at high redshift, the radiative transfer models can capture the evolution in the population more effectively than an empirical calibration based on objects in a small range of metallicities.

\section{Summary}\label{section:summary}

In this paper we present the incorporation of the \cite{Feltre2016} NLR models into \beagle. This addition allows the mixing of emission from \HII\ regions with that from the NLR of type-2 AGNs. Dust attenuation is applied self-consistently taking into account the dust within the NLR, \HII\ regions and the diffuse ISM. 

We take a pedagogical approach to defining which parameters of our model (from both \HII\ and NLR regions) can be constrained by fitting a given set of observables in the optical range. To determine this we work with idealised $z\sim0$ and $z\sim2$ galaxy spectra with varying contribution of the NLR to \Hbeta\ flux.  We fit to a set of line ratios (\OI/\Halpha, \NII/\Halpha, \SII/\Halpha, \OIII/\Hbeta, \Hbeta/\Halpha, \OI/\OII\ and \OII/\OIII) as well as the \Halpha\ line flux. We then quantify the S/N required to constrain the parameters of our model for different NLR contributions. Finally, we compare the results of our model with those obtained using previously published methods. The main results are the following:

\begin{itemize}
    \item With the set of observables used, the retrieval of physical parameters for the AGN dominated spectra with high S/N shows degeneracies between \xidAGN\ and \logUsAGN\ that lead to an underestimation of \logUsAGN\ and \ZAGN\ if \xidAGN\ is left as a free parameter. We obtain similar results when adding to the observables a ratio sensitive to the N/O abundance. Allowing \PLalpha\ to vary leads to an over-estimated retrieval of \Lacc\ while still under-estimating \ZAGN. We find the best recovery of \Lacc, \ZAGN and \logUsAGN\ when fixing both \xidAGN\ and \PLalpha.
    \item  The retrieval of physical parameters for the star-formation dominated spectra is also biased by the nebular \xid. We find that \xid\ is poorly constrained and when left free leads to an over-estimation of the \logUs\ and \Zhii, as well as tighter constraints on the biased estimates.
    \item We also test the case of equal-contributions of NLR and \HII\ regions to the \Hbeta\ flux. In this case, the accuracy of the \PLalpha\ parameter depends heavily on the region of the observable parameter space that is being probed, so we suggest that it should be fixed, unless varying it is required to reach an acceptable fit. Both \xid\ and \xidAGN\ are degenerate with several parameters when left free, biasing the recovery of these parameters. We obtain the best constraints for the objects when both \xid\ and \xidAGN\ are fixed.
    \item We have investigated the S/N in \Hbeta\ required to derive un-biased parameter estimates (while fixing \xid, \xidAGN\ and \PLalpha). AGN parameters are well constrained with a $\textrm{S/N(H}\beta)\sim20$ for the AGN-dominated galaxies. \HII\ region parameters are well constrained with $\textrm{S/N(H}\beta)\sim10$ for star-formation dominated galaxies.  However, we need a higher $\textrm{S/N(H}\beta)\sim30$ to disentangle the contributions from star-forming and NLR components when they have similar contributions to \Hbeta\ flux.
    \item We test how well \beagle\ can detect the presence of an AGN by retrieving the fractional contribution of the NLR to the total \Hbeta\ flux of a galaxy. For the $z=0$ object, we find that a S/N$\sim$10 already allows for identification of objects that have a significant contribution to the flux, even for the most challenging case of equal contributions to the flux from NLR and \HII\ regions. For the $z=2$ object, a NLR contribution is detected but overestimated at S/N$\sim$10, though we suggest this is still a reasonable limit for which to find objects with a significant NLR contribution.
    \item We have compared the predictions of our model to those from different sets of models and data compiled in the literature. These include the models from \citet{Thomas2018NebularBayes} and \citealt{Perez-Montero2019} and the calibrations to derive gas-phase oxygen abundance from \citealt{Storchi-Bergmann1998} and \citealt{Dors2021} for the data compiled by these last authors. We find that in general, all the models have a good coverage of the parameter space where the data lie. However, we find an opposite dependence of the \citealt{Perez-Montero2019} models of R23 and \OII/\OIIIall\ on \logUsAGN\ with respect to those of F16 and \citealt{Thomas2018NebularBayes}. This is presumably due to the choice made in \citealt{Perez-Montero2019} to publish only the highest \logUs\ models, as the behaviour between \logUs\ and \OII/\OIIIall\ is double-valued in their models. 
    \item To constrain the ionization state of NLRs in observed type-2 AGN we measure the \HeII\ recombination line that has a similar ionization energy to \textsc{Oiv}, in 463 confirmed type-2 AGNs from DR7 SDSS spectra. From the comparison to the data we see how the data follow the trend of the F16 models and they suggest that these data have a siginificantly lower ionization parameter values than those covered by the \citealt{Perez-Montero2019} models.
    \item To better capture the dispersion of the N/H-O/H relation in the data, we create a grid of different scaling factors of this relation that we compare to the data and the same set of models from the literature. Whereas F16 and \citealt{Perez-Montero2019} models sample well the region of the parameter space where the data lies, the fact that \citealt{Thomas2018NebularBayes} models fix the Nitrogen abundance might lead to an underestimation of the \logUsAGN\ parameter when fitting to nitrogen emission lines, because models with extremely low \logUsAGN\ values are needed to sample the data.
    \item We also compare the models to the empirical emission-line calibration from \citealt{Dors2021}.  Within the region sampled by observations, the models agree well with the empirical calibration.  However, at high metallicities, the models curve up from the plane that the empirical calibration draws in the \logOH--P--R23 space. The observations are also scattered preferentially above the plane in the region covered by the models.  Therefore, it is likely the models can capture the behavior of the observations that also scatter above the plane. At lower metallicities, the models lie below the \citealt{Dors2021} calibration.  Additionally, the difference between models with different \PLalpha\ values is less significant, which will help constraining the gas parameters in galaxies at high redshifts, such as those expected to find with missions such as the \textit{JWST}. In general, the radiative transfer models can capture the behaviour of objects with different oxygen abundances more effectively than an empirical calibration based on objects with a limited range of metallicities.
\end{itemize}

This is the first in a series of three papers, that will be followed by a paper on fitting of a sample of type-2 AGNs with \beagle, and a study of the extent to which emission from shocks and post-AGB stars may affect our inferences.

\section*{Data Availability}
The astrophysical data used in this publication is supplied within the paper and the online supplementary materials.  The idealised spectra analysed in Section 3 will be shared with reasonable request to the corresponding authors.

\section*{Statement of contributions by authors}
This is a highly collaborative project led by ECL where every author has had an important contribution. In particular for this paper, AF and ECL incorporated the models into BEAGLE. AVG and AP defined and fitted to the idealised models in Section 3 from the parameters that MH provided from the simulations. AVG worked on the discussion comparing to other models which led to ECL measuring HeII line fluxes in SDSS data and also the need of creating a new N/O grid done by AP and AF. The writing has been done by ECL with the help of AVG, and editing by SC. SC, MH and JC have contributed to the scientific discussions and followed the project from the beginning. If the field of astronomy had the mechanism for assigning multiple first authors (as is done in other fields), AVG, AP and ECL would have been assigned as lead authors.  We reflect this with the fact that they are assigned as corresponding authors.

\section*{Acknowledgments}
The authors acknowledge support from the European Research Council (ERC) via an Advanced Grant under grant agreement no. 321323-NEOGAL.
AVG also acknowledges support from the European Research Council  through the Advanced Grant MIST (FP7/2017-2020, No 742719). ECL acknowledges support of an STFC Webb Fellowship (ST/W001438/1). MH acknowledges financial support from the Swiss National Science Foundation via the PRIMA grant “From cosmic dawn to high noon: the role of BHs for young galaxies” (PROOP2$\_$193577). AF acknowledges support from grant PRIN MIUR 2017 20173ML3WW. JC acknowledges funding from
the ERC Advanced Grant 789056 “FirstGalaxies” (under the European Union’s Horizon 2020 research and innovation programme). \pyspeclines\ uses the python package \pyspeckit\ 
\url{https://pyspeckit.readthedocs.io/en/latest/index.html}

\bibliographystyle{mnras}
\bibliography{bibliography.bib}

\label{lastpage}

\end{document}